\documentclass[useAMS,usenatbib]{mn2e}
\usepackage{amssymb}
\usepackage{aas_macros}
\usepackage{natbib}
\usepackage{times}
\usepackage{graphicx}
\usepackage{color}

\newcommand{\Msun}                    {\,{\rm M}_\odot}

\newcommand{\hkpc}                     {\,h^{-1}\,{\rm kpc}}
\newcommand{\hMpc}                    {\,h^{-1}\,{\rm Mpc}}
\newcommand{\hMsun}                  {\,h^{-1}\,{\rm M}_\odot}

\newcommand{\Zsun}                     {\,{\rm Z}_\odot}

\newcommand{\erms}             {$e_\mathrm{rms}\,\,$}

\newcommand{\newl} 	{\mathrm{log_{10}}}

\def\lsim{\mathrel{\lower0.6ex\hbox{$\buildrel {\textstyle <}
 \over {\scriptstyle \sim}$}}}

\title[Galaxy-halo misalignment]{The alignment and shape of dark matter, stellar, and hot gas distributions in the EAGLE and cosmo-OWLS simulations
}%
 \author[M. Velliscig et al.]{Marco Velliscig,$^{1}$\thanks{E-mail: velliscig@strw.leidenuniv.nl} 
Marcello Cacciato,$^{1}$   
Joop Schaye,$^{1}$ 
Robert A. Crain,$^{1,2}$ \newauthor
Richard G. Bower,$^3$  
Marcel P. van Daalen,$^{1,4,5}$ 
Claudio~Dalla~Vecchia,$^{6,7}$ 
Carlos S. Frenk,$^3$ \newauthor
Michelle Furlong,$^3$ 
I.~G. McCarthy,$^2$ 
Matthieu Schaller,$^3$ 
Tom Theuns$^{3}$ 
\\\\
$^1$Leiden Observatory, Leiden University, P.O. Box 9513, 2300 RA Leiden, The Netherlands \\
$^2$Astrophysics Research Institute, Liverpool John Moores University, 146 Brownlow Hill, Liverpool L3 5RF \\
$^3$ Institute for Computational Cosmology, Department of Physics, University of Durham, South Road, Durham, DH1 3LE, UK\\
$^4$Max Planck Institute for Astrophysics, Karl-Schwarzschild Stra\ss{}e 1, 85741 Garching, Germany\\
$^5$Department of Astronomy, Theoretical Astrophysics Center, and Lawrence Berkeley National Laboratory, University of California, Berkeley, CA 94720, USA\\
$^6$Instituto de Astrof\'isica de Canarias, C/ V\'ia L\'actea s/n, 38205 La Laguna, Tenerife, Spain\\
$^7$Departamento de Astrof\'sica, Universidad de La Laguna, Av. del Astrof\'isico Franciso S\'anchez s/n, 38205 La Laguna, Tenerife, Spain \\
}

\begin{document}

\date{\today}
\pagerange{\pageref{firstpage}--\pageref{lastpage}} \pubyear{2014}

\maketitle

\label{firstpage}
\begin{abstract}
We report the alignment and shape of dark matter, stellar, and hot gas distributions in the EAGLE and cosmo-OWLS simulations.
The combination of these state-of-the-art hydrodynamical cosmological simulations enables us to span four orders of magnitude in halo mass ($11 \le \newl({M_{200}/ [\hMsun]}) \le 15$), a wide radial range ($-2.3 \le \newl(r/[\hMpc]) \le 1.3$) and redshifts $0\le z\le 1$.
The shape parameters of the dark matter, stellar and hot gas distributions follow qualitatively similar trends: they become more aspherical (and triaxial) with increasing halo mass, radius and redshift. 
We measure the misalignment of the baryonic components (hot gas and stars) of galaxies with their host halo as a function of halo mass, radius, redshift, and galaxy type (centrals {\it vs} satellites and early- {\it vs} late-type). 
Overall, galaxies align well with the local distribution of the total (mostly dark) matter.
However, the stellar distributions on galactic scales exhibit a median misalignment of about 45-50 degrees with respect to their host haloes. This misalignment is reduced to 25-30 degrees in the most massive haloes ($13 \le \newl(M_{200}/ [\hMsun]) \le 15$). 
Half of the \emph{disc} galaxies in the EAGLE simulations have a misalignment angle with respect to their host haloes larger than 40 degrees.
We present fitting functions and tabulated values for the probability distribution of galaxy-halo misalignment to enable a straightforward inclusion of our results into models of galaxy formations based on purely collisionless N-body simulations. 

\end{abstract}
\begin{keywords}
cosmology: large-scale structure of the Universe, cosmology: theory, galaxies: haloes, galaxies: formation
\end{keywords}


\section{Introduction}
\label{Sec:introduction}

The topology of the matter distribution in the Universe is well described as a web-like structure
comprising voids, sheets, filaments and haloes. 
This so-called {\it cosmic web} arises naturally from the gravitational growth
of small initial perturbations in the density field of an expanding cold dark matter dominated ($\Lambda$CDM) Universe.
The evolution of the properties of the large-scale cosmic web is governed by the dominant components, i.e. dark energy and dark matter, 
while baryons are expected to trace the distribution of the latter. 
Specifically, galaxies reside in dark matter haloes and trace them in terms of their positions and, to first order, in terms of their shapes and mutual alignment, albeit in a biased fashion due to the dissipative processes they experience during galaxy formation. 
Theoretical studies of this {\it galaxy bias} have been ongoing for several decades \citep[e.g.][]{Kaiser84,Davis85}.

It has become apparent that when galaxies are used to infer the properties of the underlying 
dark matter distribution, it is convenient to bisect this investigation into two steps: 
the relation between galaxies and haloes and the relation between haloes and the underlying density field. 
The latter can be studied directly via cosmological N-body simulations, whereas the former is a far more complicated relation that is potentially affected by virtually all the physical processes associated with galaxy formation. For instance, while the triaxial shape of dark matter haloes is understood in terms of the collisionless nature of dark matter coupled with ellipsoidal collapse, galaxies manifest 
themselves in a plethora of morphologies ranging from thin to bulge-dominated discs and to ellipsoidals and this is undoubtedly related to the redistribution of angular momentum occurring during galaxy 
formation and evolution which, in turn, depends on the physical processes in operation. Thus, the characterization of the way galaxy shapes relate to their host haloes holds the potential to unveil the relevant physical mechanisms behind such a rich manifestation of galaxy types.

Numerical simulations have been used to study the mutual alignment of 
galaxies with their own host haloes. For instance, 
\citet[][]{vandenBosch02}, \citet[][]{Chen03}, \citet[][]{Sharma05}, \citet[][]{Bett10} and \citet[][]{Sales12}
have shown that the angular momentum distributions of gas and dark matter components
are partially aligned, with a typical misalignment angle
of $\sim 30^{\circ}$, although this might predominantly apply
to disc galaxies. On the other hand, central ellipticals are expected to be aligned with their host haloes if they are formed by mergers \citep[][]{Dubinski98, Naab06, Boylan-Kolchin06}, because the orientations of the central ellipticals
and of the host dark matter haloes are determined by respectively the orbital angular momenta of their  (correlated) progenitor galaxies and haloes. 
Observationally, there exist different indications of the presence of a misalignment between galaxies and their host haloes. 
However, different studies have reached somewhat conflicting conclusions about the typical values of this misalignment angle \citep[see e.g.][]{Heymans04, Kang07, Wang08, Okumura09}.

Beyond its theoretical relevance, the misalignment of a galaxy with its own host halo can be a source of systematics for those studies that aim to: infer the shape of dark matter haloes or constrain cosmological parameters via the measurement of the galaxy shape correlation function.  
Several current and forthcoming weak lensing surveys (e.g. KiDS, DES, LSST, and Euclid\footnote{KIDS:
  KIlo-Degree Survey, \\ http://www.astro-wise.org/projects/KIDS/; \\
  DES: Dark Energy Survey, https://www.darkenergysurvey.org;\\ LSST:
  Large Synoptic Survey Telescope, http://www.lsst.org; \\ Euclid:
  http://www.euclid-ec.org})
 will achieve the statistical power to probe, observationally, halo shapes and to obtain exquisite measurements of the apparent alignment of galaxy shapes --\emph{cosmic shear}-- due to the gravitational lensing effect caused by the underlying (dark) matter distribution. It is therefore of great importance to guide the interpretation of the measured signal with numerical simulations. For instance, the link between the shape of the visible, baryonic matter and the structure of the underlying dark matter distribution, as well as their mutual orientation can be examined. 
To this end it is necessary to complement the expectations derived from cosmological N-body simulations
with the properties of galaxies as inferred from small-scale, high-resolution hydrodynamic simulations and/or semi-analytical models \citep[e.g.][]{Joachimi13, vandenBosch02, Croft09, Hahn10, Bett10, Bett12}.

In this paper, we extend previous work by exploiting the wealth of information encoded in hydro-cosmological simulations
in which the main physical processes responsible for galaxy formation and evolution are simultaneously at play,
thus leading to a more realistic realization of the galaxy-dark matter connection.
We use  the OverWhelmingly Large Simulations 
 \citep[cosmo-OWLS,][]{Schaye10,LeBrun14,McCarthy14}
and the Evolution and Assembly of GaLaxies and their  Environments \citep[EAGLE,][]{Schaye14,Crain15} project.
This approach has the advantage that the processes that lead to galaxy formation are self-consistently incorporated in the simulations
and are therefore accounted for in the resulting galaxy and halo shapes, as well as in their correlation.
During the late phase of this project, a study adopting a similar methodology was \citet{Tenneti14a}, hereafter Ten14, which has many aspects in common  with our analysis. Throughout the paper, 
we will therefore compare mutual findings.

Our study is, however, unique as a consequence of several key features of our simulations and analysis.
As detailed in \S~\ref{Sec:simulations}, the use of cosmo-OWLS and EAGLE provides us sufficient 
cosmological volume and resolution, both of which are crucial for the reliability and the applicability of our results. 
Specifically, we span four orders of magnitude in halo mass ($11 \le \log(M_{200}/[\hMsun]) \le 15$) and over six orders of magnitude in subhalo mass $M_{sub}$, enabling us to investigate spatial variations of the shape of galaxies and haloes from galactic to cosmological scales.
Furthermore, the combination of EAGLE and cosmo-OWLS forms a set of simulations that reproduces the observed abundance of galaxies as a function of stellar mass
(the galaxy stellar mass function) at both low ($\log(M_{200}/[\hMsun]) \le 13$) and high ($13 \le \log([M_{200}/[\hMsun]) \le 15$) halo masses. Moreover, it has been shown that the cosmo-OWLS simulations reproduce various (X-ray and optical) observed properties of galaxy groups \citep{Crain10,McCarthy10,LeBrun14} as well as the observed galaxy mass function for haloes more massive than $\log(M_{200}/[\hMsun]) = 13$. Finally, the galaxy size distribution in EAGLE reproduces the observed one \citep{Schaye14}.

This paper is organized as follows. We summarize the properties of the simulations in \S~\ref{Sec:methods}, where we also introduce the technical definitions used throughout the paper. 
We highlight some caveats to the shape and angle estimates related to the feedback implementation in \S~\ref{Sec:CaveatsGalForm}.
In \S~\ref{Sec:ShapeParametersResults} we present the results concerning the sphericity and triaxiality of dark matter haloes, as well as those of the stellar and the hot X-ray emitting gas distribution. 
The (mis)alignment of the baryonic components with their host haloes is addressed in \S~\ref{Sec:MisAlignment}.
We summarize and comment on our results in \S~\ref{Sec:Conclusions}.

Throughout the paper, we assume a flat $\Lambda$CDM cosmology with massless neutrinos. 
Such a cosmological model is characterized by five\footnote{Flatness implies that $\Omega_{\Lambda} = 1-\Omega_{ \rm m}$.} parameters: 
\{$\Omega_{ \rm m}$, $\Omega_{ \rm b}$, $\sigma_{ \rm 8}$, $n_{ \rm s}$, $h$\}. 
The simulations used in this paper were run with two slightly different sets of values for these parameters.
Specifically, we will refer to PLANCK as the set of cosmological values 
suggested by the Planck mission \{$\Omega_{ \rm m}$,\, $\Omega_{ \rm b}$,\,$\sigma_{ \rm 8}$,\, $n_{ \rm s}$,\, $h$\} = \{0.307, 0.04825, 0.8288, 0.9611, 0.6777\} (Table 9; \citealt{Planck13}), whereas WMAP7 refers to the cosmological parameters \{$\Omega_{ \rm m}$, $\Omega_{ \rm b}$, $\sigma_{ \rm 8}$, $n_{ \rm s}$, $h$\} = \{0.272, 0.0455, 0.728, 0.81, 0.967, 0.704\} suggested by the 7th-year data release \citep{wmap7} of the WMAP mission. 

\section{Simulations and Technical Definitions}
\label{Sec:methods}

\begin{table*}
\begin{minipage}{168mm}
\begin{center}
\caption{Simulations used throughout the paper and their relevant properties. Description of the columns: (1) descriptive simulation name; (2) comoving size of the simulation volume; (3) total number of particles; (4) cosmological parameters used in the simulation; (5) initial mass of baryonic particles; (6) mass of dark matter particles; (7) maximum softening length; (8) colour used for the simulation; (9) simulation name tag.} 
\begin{tabular}{lllccccll}
\hline
Simulation  & L & $N_{\rm particle}$& Cosmology & $m_{\mathrm{b}}  $ & $m_{\mathrm{dm}}$ & $\epsilon_{\rm prop}$ & Colour & tag \\
 & & &  & $ [\hMsun] $ & $[\hMsun]$ & $(\hkpc)$ &  &  \\
 (1)&(2)& (3)& (4)& (5)& (6)& (7)&(8) & (9)\\ 
\hline 
{EAGLE Recal} &  25 $(\rm Mpc)$   & $2 \times 752^3$  & PLANCK & $ 1.5 \times 10^5$  & $ 8.2 \times 10^5$ & $0.5$ & purple     & EA L025\\
{EAGLE Ref}    & 100 $(\rm Mpc)$   & $2 \times 1504^3$ & PLANCK & $ 1.2 \times 10^6$  & $ 6.6  \times 10^6$ & $0.2$ & orange \footnote{Cyan is used for Figs.~\ref{fig:galtype_star_vs_halo_MinR_orient} and  \ref{fig:satcen_star_vs_halo_MinR_orient} where the EA L100 simulation is used in order to improve the statistics for the least massive bin.}      & EA L100\\
{cosmo-OWLS AGN 8.0}      & 200 $(\hMpc)$ & $2 \times 1024^3$ & WMAP7  & $ 8.7  \times 10^7$  & $ 4.1  \times 10^8$ & $2.0$ &  blue & CO L200\\
{cosmo-OWLS AGN 8.0}      & 400 $(\hMpc)$ & $2 \times 1024^3$ & WMAP7  & $ 7.5  \times 10^8$  & $ 3.7  \times 10^9$ & $4.0$ &  green & CO L400\\
\hline
\end{tabular}

\label{tbl:sims} 
\end{center}
\vspace{-0.3in}
\end{minipage}
\end{table*}

\subsection{Simulations}\label{Sec:simulations}

Throughout the paper, we employ the outputs of four cosmological volumes simulated within the context of two distinct projects: EAGLE \citep{Schaye14, Crain15} and cosmo-OWLS \citep{LeBrun14, McCarthy14}. We use the former to investigate (well-resolved) smaller halo masses in relatively small volumes; whereas the latter is used to study more massive haloes in larger volumes. Table~\ref{tbl:sims} lists all relevant specifics of these simulations.

Both EAGLE and cosmo-OWLS were run using a modified version of the $N$-Body Tree-PM smoothed particle hydrodynamics (SPH) code \textsc{gadget}~3, which was last described in \citet{Springel05_gadget}. The main modifications are the formulation of the hydrodynamics, the time stepping and, most importantly, the subgrid physics. All the simulations used in this work include element-by-element radiative cooling (for 11 elements; \citealt{Wiersma09a}), star formation \citep[][]{Schaye08}, stellar mass loss \citep{Wiersma09b}, energy feedback from star formation \citep{DallaVecchia08,DallaVecchia12}, gas accretion onto and mergers of supermassive black holes (BHs; \citealt{Booth09, Rosas13}), and AGN feedback \citep{Booth09, Schaye14}. 

The subgrid physics used in EAGLE builds upon that of OWLS 
 \citep[][]{Schaye10}, GIMIC \citep{Crain09} and cosmo-OWLS 
 \citep{LeBrun14,McCarthy14}. Furthermore, the EAGLE project brings a number of changes with respect to cosmo-OWLS regarding the implementations of energy feedback from star formation (which is now thermal rather than kinetic), the accretion of gas onto BHs (which now accounts for angular momentum), and the star formation law (which now depends on metallicity). 
More information regarding technical implementation of hydro-dynamical aspects as well as subgrid physics can be found in \citet{Schaye14}.

Arguably, the most important feature of the EAGLE simulation is the calibration of the subgrid physics parameters to reproduce the observed galaxy mass function and galaxy sizes at redshift zero.
One of the key feature of the cosmo-OWLS simulations is that they reproduce optical and X-ray scaling relations of groups and clusters of galaxies.
In this work we exploit both these unique features by splitting our range of halo masses into four mass bins and by using a different simulation for each one of them. Specifically, for halo masses below the `knee' of the galaxy stellar mass function we use EAGLE in order to ensure galaxies form with the `correct' efficiency and size, whereas for haloes above the `knee' we use cosmo-OWLS. In practice, we create a composite sample of haloes spanning four orders of magnitude in mass ($11 \le \log(M_{200}/ [\hMsun]) \le 15$).

\begin{table*} 
\begin{minipage}{168mm}
\begin{center}
\caption{Values at $z=0$ of various quantities of interest in each mass bin. Description of the columns: (1) simulation tag; (2) mass range $\newl({M}_{\rm 200}/[\hMsun])$ of the haloes selected from the simulation; (3) median value of the halo mass $\newl(M_{\rm 200}^{\rm crit})$; (4) median value of the stellar mass ($\newl({M}_{\rm star}/[\hMsun])$) considering all the star particles that belong to the halo; (5) standard deviation of the stellar mass distribution $\sigma_{\newl  M_{\rm star}} $; (6) median value of halo radius $r_{200}^{\rm crit}$; (7) median radius within which half of the mass in dark matter is enclosed; (8) median radius within which half of the mass in stars is enclosed; (9) number of haloes; (10) number of {\emph{ satellite} haloes}.} 
\begin{tabular}{lccccrrrrr}
\hline
Simulation tag &  mass bin & $M_{\rm 200}^{\rm crit}$ & $M_{\rm star} $ & $\sigma_{\newl  M_{\rm star}} $ & $r_{200}^{\rm crit}$ & $r_{\rm half}^{\rm dm} $ & $r_{\rm half}^{\rm star} $ & $N_{\rm halo}$ & $N_{\rm sat}$ \\
 & * & * & * & * & ** & ** & ** &  &  \\
 (1)   &(2)            & (3)     & (4)    & (5)    & (6)    & (7)    &(8)  &(9)  &(10)\\
\hline
EA L025 & $[11-12]$ & $11.31$ & $9.50$ & $0.45$ & $96.0$ & $39.8$ & $2.7
$ & $156$ & $24$ \\
EA L100 & $[12-13]$ & $12.27$ & $10.58$ & $0.26$ & $199.3$ & $93.4$ & $4.9
$ & $1008$ & $104$ \\
CO L200 & $[13-14]$ & $13.16$ & $11.21$ & $0.25$ & $396.4$ & $241.8$ & $
53.4$ & $2190$ & $137$ \\
CO L400 & $[14-15]$ & $14.09$ & $12.06$ & $0.19$ & $805.9$ & $505.1$ & $
106.7$ & $1152$ & $26$ \\
\hline                                                   
\end{tabular}
\label{tbl:sims_stat} 
\end{center}
\vspace{-0.2in}
\footnotetext{* $\newl (M/[\hMsun])$}
\footnotetext{** $R/[\hkpc]$}
\end{minipage}
\end{table*}

\subsection{Halo and subhalo definition}
\label{Sec:methods_halodef}
Groups of particles are identified in our simulations by applying the Friends-of-Friends algorithm with linking length $0.2$ to the dark matter particles 
 \citep{Davis85}. The mass $M_{200}^{\rm crit}$ and the radius $r_{200}^{\rm crit}$ of the groups are assigned using a spherical over-density algorithm centred on the minimum of the gravitational potential, as implemented in \textsc{subfind} 
  \citep{Springel01_subfind,Dolag09}. 
From each group, dynamically un-bounded particles are discarded. { Thus, subhaloes are identified as a collection of bound particles that reside in a local minimum of the gravitational potential computed using all particle types.} The most massive subhalo is the \emph{central} subhalo of a given FoF group and all other subhaloes are \emph{satellites}.
Particles that are bound to a subhalo belong exclusively to that subhalo. Correspondingly, central subhaloes do not contain particles that reside in other local minima of the potential, even if those particles are within the subhalo boundary. We define the centre of a subhalo as the position of the particle with the minimal gravitational potential. The subhalo radius can be calculated for each component separately. A commonly used estimate is the radius within which half of the mass in dark matter is included, $r_{\rm half}^{\rm dm}$. The mass of a subhalo is the sum of the masses of all the particles that constitute it.
For the rest of the paper we will use the term `halo' to refer both to central and satellite subhaloes, unless otherwise specified.

The masses of subhaloes for both centrals and satellites (according to \textsc{subfind} classifications), are indicated with $M_{\rm sub}$. However, whenever a distinction is required, we shall use $M_{200}$ and $r_{200}$ to characterize the properties of central haloes. 

\subsection{Shape parameter definitions}
\label{Sec:ShapeParameters}
 
A fundamental quantity that describes how matter is spatially distributed  is the three-dimensional mass distribution tensor \citep[ e.g.][]{Davis85, Cole96}, 
\begin{equation}\label{eq:inertiatensor}
M_{ij}= \sum_{p=1}^{N_{\rm P}} m_p x_{pi} x_{pj} \, ,
\end{equation}
where $N_{\rm P}$ is the number of all particles that belong to the structure of interest, 
$x_{pi}$ denotes the element $i$ (with $i,j=1,2,3$ for a 3D particle distribution) of the position vector of particle $p$, and $m_p$ is the mass of the $p^{th}$ particle.
This mass distribution tensor is often referred as the inertia tensor, since the two tensors share the same eigenvectors \citep[see][for a discussion]{Zemp11} and,  
for most astrophysical purposes, those eigenvectors encode the information of interest. 
Throughout this paper we will refer to the mass distribution tensor as the inertia tensor 
to conform to the jargon used in the literature.  

The eigenvalues of the inertia tensor will be denoted as 
$\lambda_i$ (with $i=1,2,3$ for a 3D particle distribution as in our case). 
Given a particle distribution inertia tensor, 
the modulus of the major, intermediate, and minor axes of the corresponding ellipsoid can be written
in terms of these eigenvalues as 
$a=\sqrt{\lambda_1}$, $b=\sqrt{\lambda_2}$, and $c=\sqrt{\lambda_3}$, respectively.
We interpret this ellipsoid as an approximation to the shape of the halo. 
Specifically, the sphericity and triaxiality parameters, $S$ and $T$, are defined as
\begin{eqnarray}\label{eq:shapeparameters}
S = \frac{c}{a} \, ,  \, \, \, {\rm and} \, \, \,
T = \frac{a^2 - b^2 }{a^2 - c^2 } . 
\end{eqnarray}
A purely spherical halo will have $S = 1$ with $T$ being undefined. Low values of $T$  (i.e. $T \rightarrow 0$) 
correspond to oblate haloes while high values (i.e. $T \rightarrow 1$) correspond to prolate haloes.

We note that the computation of shape parameters in a spherical region biases the shape towards higher sphericity. When computing shapes of dark matter haloes in spherical regions it is possible to correct for this effect applying the simple empirical re-scaling: $S_{true}=S^{\sqrt{3}}$ as suggested in \citet{Bailin05}. This correction is not implemented in the results presented here since a similar correction is not available for the other quantities that we present.

\begin{figure}
\begin{center} 
\includegraphics[width=1.0\columnwidth]{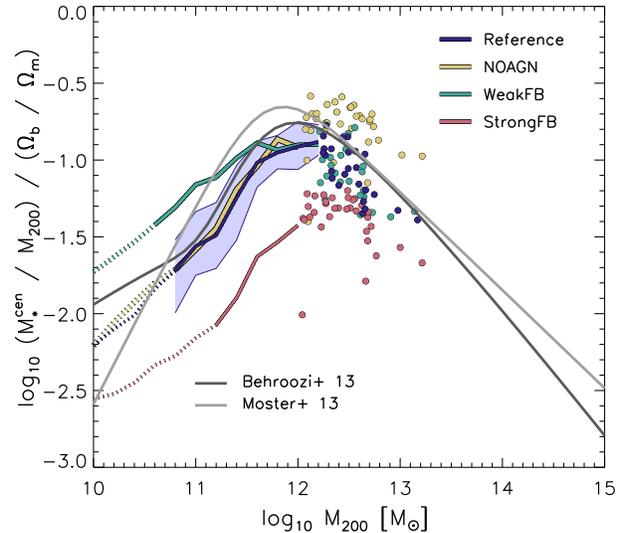} 
\end{center}
\caption{ {The stellar mass to halo mass ratio of central galaxies as a function of halo mass, normalised by the cosmic baryon fraction, for the four feedback variations used in \S\ref{Sec:CaveatsGalForm}. The curves are dotted where there are fewer than 100 star particles per galaxy and individual galaxies are showed for bins that contain fewer than 10 galaxies. The 1-sigma scatter about the median of Reference is shown as a shaded region. Dark and light grey lines represent the abundance matching relations of \citet{Behroozi13} and \citet{Moster13}.} 
}
\label{fig:L025_galeff}
\end{figure}

\begin{figure*}
\begin{center} 
\begin{tabular}{cc}
\includegraphics[width=1.0\columnwidth]{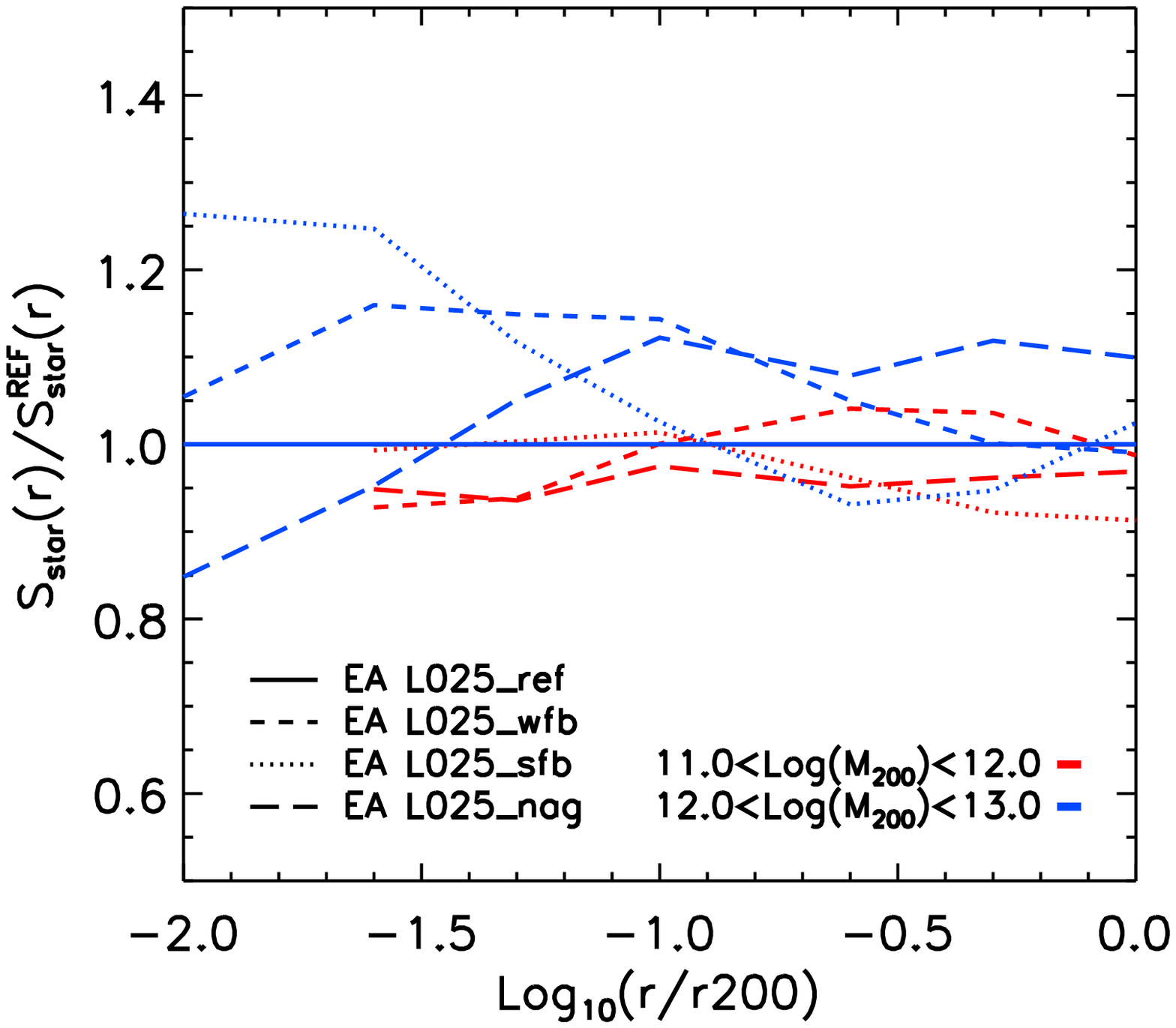} &
{\includegraphics[width=1.0\columnwidth]{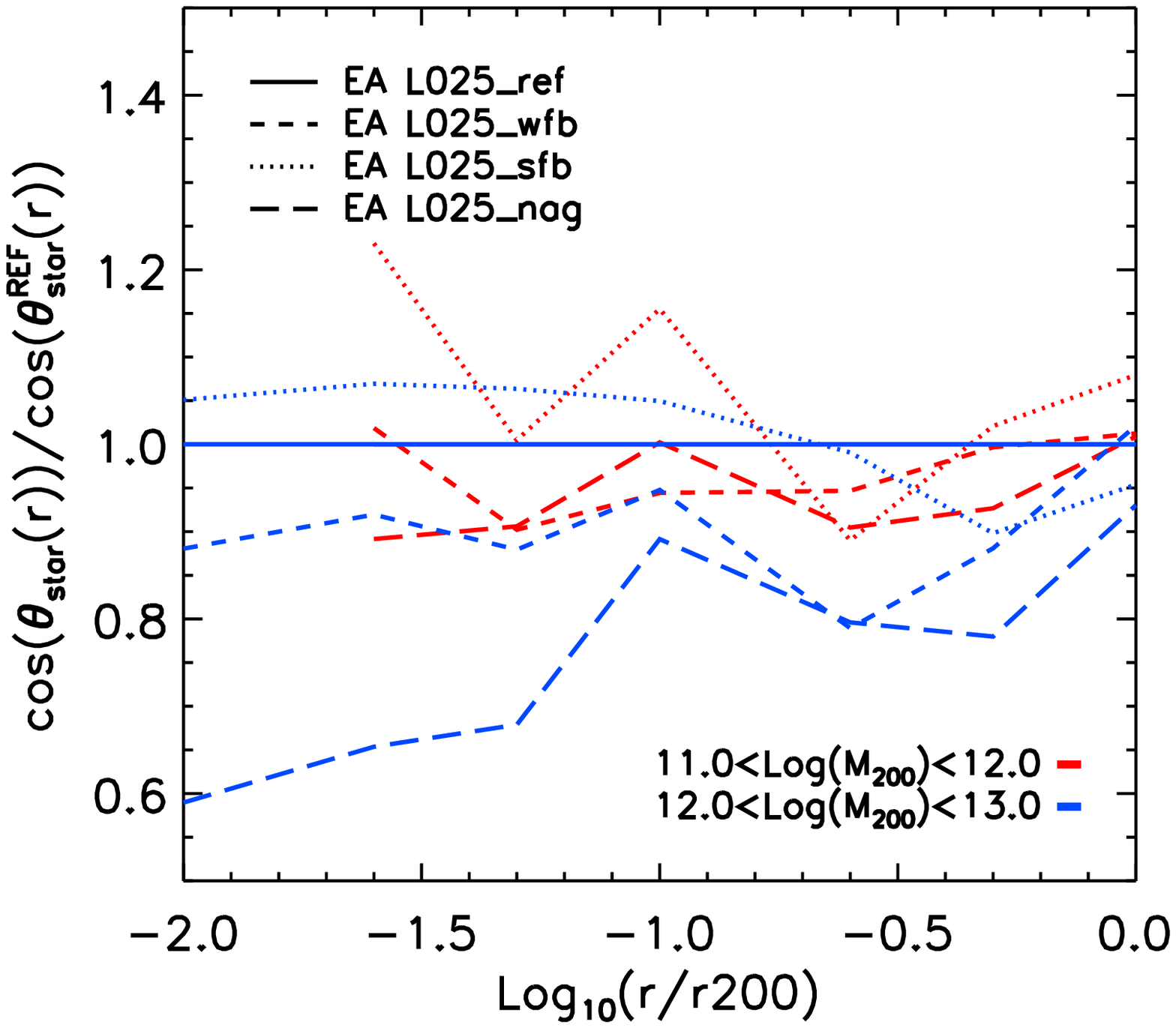}} \\
\end{tabular} 
\end{center}
\caption{ 
Ratios of the average sphericity (left panel) and average misalignment angles (right panel) of the stellar distribution with respect to the average values of the REFERENCE EAGLE simulation (see \S\ref{Sec:CaveatsGalForm}). Different colours indicate different mass bins while different line styles refer to different simulations which differ only by the implementation of feedback.
The disagreement (up to 20\% for sphericity and up to 40\% for misalignment angle) stems from the different efficiency of galaxy formation (see discussion in \S~\ref{Sec:CaveatsGalForm}). The misalignment angle is more sensitive to galaxy formation efficiencies than the sphericity. The differences always increase towards the centre of the halo.
}
\label{fig:L025_galeff_test}
\end{figure*} 

\subsection{Axes and misalignment angle definition}
\label{Sec:MisAlignmentDefinition}
The eigenvectors of the inertia tensor, in Eq.\ref{eq:inertiatensor}, are denoted as ${\bf e}^{\,\,i}_{x}$, with $i=1,2,3$ in the case of a 3D distribution of particles and $x= \rm halo,star,gas$ to indicate total matter\footnote{We do not deal with the specific case of only dark matter because on the scales of interest it almost exactly coincides with the total matter in a halo.}, stars, or gas, respectively. 
We relate the ordered eigenvectors ${\bf e}^{\,\,1}_{x}$, ${\bf e}^{\,\,2}_{x}$, and ${\bf e}^{\,\,3}_{x}$ of the inertia tensor to the direction of the major, intermediate and minor axis of the corresponding ellipsoid. 
We further indicate the radial dependence of the major axis as $e^1_{x}(r)$, which, unless stated otherwise, has been computed using the volume enclosed by the entire structure as defined by SUBFIND (see \S\ref{Sec:methods_halodef}). 
We shall quantify the alignment of different matter components via the scalar product of two major axes, i.e. the misalignment angle $\theta$ ($\Theta$ in case of projected quantities). Specifically, we will use $\cos{\theta}$ as the principal quantity of interest and only comment on the actual value of $\theta$ when relevant. 
We stress here that the major axis is a spin-2 quantity, i.e. it is invariant under rotation of 180 degrees. This means that $\theta$ only varies between 0 and 90 degrees and, correspondingly, $\cos{\theta}$ can only assume values between zero and unity.

\section{The effect of galaxy formation efficiency}
\label{Sec:CaveatsGalForm}

A major asset of our composite sample of simulated haloes is that it reproduces the observed stellar-to-halo mass ratio as a function of halo mass. Specifically, EAGLE has been calibrated to reproduce the stellar mass function at redshift zero and cosmo-OWLS has proven successful in reproducing many observable properties of groups and clusters \citep[][]{ McCarthy10, LeBrun14}. Moreover, in the halo mass range where cosmo-OWLS haloes are used, their galaxy formation efficiency is consistent with the results of \citet{Moster13,Behroozi13} from abundance matching techniques. 
This feature is particularly important in the context of our investigation, as one might expect that if a simulation produces either too many or too few stars, then their distribution and consequently, the galaxy shape parameters would also be affected. Note however that, as shown by \citet{Crain15}, this criteria is insufficient to guarantee that the spatial distribution of barionic matter is realistic.

Before showing the main results of our analysis we investigate how different feedback implementations results in different predictions for the shape and orientations of galaxies with respect to their host haloes.
To quantify this effect, we make use of a set of feedback variations on the Reference model of the EAGLE simulations that, unlike the Reference model itself, do not reproduce the observed galaxy stellar mass function (i.e. in those simulations haloes do not form stars with the observed efficiency). A detailed description of these simulations can be found in \citet{Crain15}. Here, we only briefly summarize their properties. All simulations adopt the PLANCK cosmology. The simulation boxes have comoving volumes of $25^3 \rm{Mpc}^3$, with $2 \rm{x} 376^3$ particles. We consider four variations: 
\begin{description}
  \item[\bf{L025\_ref}] A simulation run in a smaller volume with respect to the main run using the Reference EAGLE implementations, namely: thermal energy feedback from star formation, BH gas accretion that takes into account the gas angular momentum and a star formation law which depends on gas pressure and metallicity. In the thermal feedback implementation the amount of energy injected per feedback event is fixed but there is freedom in the amount of energy that can be injected per unit of stellar mass. { This freedom is incorporated in the parameter $f_{\rm th}$ that is the expectation value of the amount of energy injected per unit stellar mass formed, in units of the energy available from core collapse supernova for our IMF. The average number of neighbouring particles heated by a feedback event is $<N_{\rm heat}>\approx1.3f_{\rm th}\left(\frac{\Delta T}{10^{7.5} K } \right)^{-1}$ whereas the temperature jump for the single particle is fixed to $\Delta T=10^{7.5} K$.  If the value of $f_{\rm th}$ is constant, then both the energy injected per single event of feedback and the energy per unit of stellar mass are fixed. By varying the parameter $f_{\rm th}$, it is possible to control the efficiency of the feedback and so to account for the unresolved radiative losses that depend on the physical state of the ISM, or to compensate for numerical losses (see \citealt{Schaye14} and \citealt{Crain15} for a discussion). The value of $f_{\rm th}$ depends on the local physical conditions (density and metallicity) of the gas according to:} 
\begin{equation}
f_{\rm th} = f_{\rm th,min} + \frac{f_{\rm th,max} - f_{\rm th,min}}
{1 + \left (\frac{Z}{0.1\Zsun}\right )^{n_Z} \left (\frac{n_{\rm H,birth}}{n_{{\rm H},0}}\right )^{-n_n}},
\label{eq:f(Z,n)}
\end{equation} 
{ where $n_{\rm H,birth}$  is the density of the parent gas particle at the time it was converted into a stellar particle and $Z$ is the gas metallicity.
The value of $n_{{\rm H},0} = 0.67~{\rm cm}^{-3}$ was chosen to reproduce the observed present-day GSMF and galaxy sizes, whereas $n_Z = n_n = 2/{\rm ln}10$. 
We use the asymptotic values $f_{\rm th,max}=3$ and $f_{\rm th,min}=0.3$, where the high asymptote $f_{\rm th,max}$ is reached at low metallicity and high density.}
  \item[\bf{L025\_wfb}] Weaker stellar feedback than for the Reference model. In this case the function in Eq.~\ref{eq:f(Z,n)} is scaled by a factor of 0.5.
  \item[\bf{L025\_sfb}] Stronger stellar feedback than for the Reference model.  In this case the function in Eq.~\ref{eq:f(Z,n)} is scaled by a factor of 2.
    \item[\bf{L025\_nag}] Same as Reference but without AGN feedback.
\end{description}
{ Fig.~\ref{fig:L025_galeff} shows the stellar mass to halo mass ratio of central galaxies as a function of halo mass, normalized by the cosmic baryon fraction, for the four aforementioned feedback variations. The galaxy stellar mass function and the galaxy sizes as obtained from these different feedback variations can be seen in Fig.~10, panels a and c, of \citet{Crain15}. Those models produce stellar mass functions with differences of the order of half a dex above (L025\_wfb) and below (L025\_sfb) the Reference one. The case without AGN feedback differs from the Reference case only for the most massive galaxies. Dark and light grey lines represent the abundance matching relations of \citet{Behroozi13} and \citet{Moster13}, respectively. The Reference simulation shows good agreement with the abundance matching models. } 

Fig.~\ref{fig:L025_galeff_test} shows the changes in the main quantities of interest in our analysis for the aforementioned feedback implementations. 
The left panel displays the ratio of the sphericity of the stellar component of haloes as a function of the distance from the halo centre for each simulation with respect to L025\_ref.
Different line styles refer to different simulations
and we report the results for two halo mass bins.
The difference  are of the order of 10\%. For the triaxiality parameter (not shown here) the differences range from 15\% to 50 \%. The right panel displays the ratio of the cosine of $\theta(r)$ (the angle between the halo's first eigenvector and the first eigenvector of the stars inside a given radius) of each simulation with respect to L025\_ref. This quantity shows 10\% differences at $r_{200}^{\rm crit}$, while differences as large as 40\% for the case without AGN (and 20\% in the case of weak SN feedback) are present on scales representative of typical galaxy sizes.
{ We report that the differences between the sphericity of haloes in the different sub-grid implementations (not shown) are smaller than 5\% at all radii.}
This analysis underlines the importance of the calibration of feedback, especially for the shape and alignment of the innermost parts of haloes where most of the stars reside.

A priori, there is no guarantee that reproducing the galaxy stellar mass function is a sufficient condition to predict realistic shape parameters. For instance, one may envision a scenario in which the size of galaxies, at the same mass,  will also influence their shapes. \citet{Crain15} have reported four different simulations in which the galaxy stellar mass function is equally well reproduced but the predictions for galaxy sizes are widely different. 
We computed the shape parameters and star-halo misalignment for the same simulations employed in \citet{Crain15}. Although in rough agreement, the relative variance from model to model is $\sim 10\% - 15\%$ for both the sphericity and the misalignment angle (not shown). Clearly, beyond the effect of the `galaxy formation efficiency', galaxy sizes also play a role in the accuracy of the retrieved shape parameters.

{ In this section and in the rest of this paper we will not focus on the origin of the different shapes and misalignment of the different populations of haloes. Investigating the physical origin of shapes and misalignments represents an interesting line of inquiry that has been addressed using zoom-in simulations and by following the evolution of galaxies and haloes in time \citep[e.g.][]{Romano-Diaz09, Scannapieco09, Cen14}.
In this work we focus on exploiting the large dynamical range available to give statistical trends with halo mass and radius and postpone a detailed investigation on their physical origin to future work. }

\section{Shape of the different components of haloes}
\label{Sec:ShapeParametersResults}

\begin{table*} 
\begin{minipage}{168mm}
\begin{center}
\caption{Values of main quantities of interest for each halo mass bin. Values refer to $z=0$ and are measured at the half-mass radius in star, $r_{\rm half}^{\rm star}$, for all subhaloes. Angle $\theta$ refers to 3D quantities, whereas $\Theta$ refers to the 2D projected equivalent. Description of the columns:(1) simulation tag; (2) mass range of the haloes, $\newl M_{200}$; (3) Median value of the subhalo mass, $\newl M_{sub}$, considering the sum of all the masses of the particles belonging to the subhalo; (4) median value of the stellar mass considering all the star particles belonging to the halo; (5) median value of the sphericity computed at the stellar half-mass radius; (6) median value of the triaxiality computed at the stellar half mass radius; (7) median value of the projected ellipticity (averaged over the three axis projections x, y and z); (8) median angle between the first eigenvector of the stellar component enclosed in $r_{\rm half}^{\rm star}$ and the first eigenvector of the total matter distribution in the halo; (9) same as (8) but for the projected haloes averaged over the three projection axes; (10) median angle between the first eigenvector of the stellar distribution and the total matter distribution, both evaluated at $r_{\rm half}^{\rm star}$; (11) same as (10) but for the projected haloes averaged over the three projection axes.} 
\begin{tabular}{lcccccccccc}
\hline
Sim &  mass bin & $ M_{\rm sub}$ & $ M_{\rm star} $ & $S $
 & $T$ & $E_{\rm 2D} $ & $\theta_{\rm halo}^{\rm star} $
 & $\Theta_{\rm halo}^{\rm star} $ & $\theta_{\rm mass}^{\rm star} $ & 
$\Theta_{\rm mass}^{\rm star} $ \\
tag & * & * & * &  &  &  & Deg & Deg & Deg & Deg \\
  (1)   &(2)            & (3)     & (4)    & (5)    & (6)    & (7)    &(8)  &(9)  &(10) &(11) \\
\hline
\vspace{0.05in}
EA L025 & $[11-12]$ & $11.33$ & $9.50$ & $0.61^{+0.17}_{-0.10}$ & $
0.22^{+0.39}_{-0.17}$ & $0.82^{+0.07}_{-0.07}$ & $47.90^{+29.60}_{-24.75}$ & $
32.44^{+18.13}_{-17.02}$ & $8.21^{+36.35}_{-6.19}$ & $4.95^{+7.67}_{-3.42}$ \\
\vspace{0.05in}
 EA L100 & $[12-13]$ & $12.28$ & $10.58$ & $0.58^{+0.11}_{-0.12}$ & $
0.31^{+0.43}_{-0.23}$ & $0.79^{+0.06}_{-0.07}$ & $46.59^{+29.75}_{-27.43}$ & $
32.34^{+18.72}_{-15.88}$ & $3.86^{+14.37}_{-2.67}$ & $3.17^{+4.99}_{-1.66}$ \\
\vspace{0.05in}
CO L200 & $[13-14]$ & $13.25$ & $11.21$ & $0.65^{+0.09}_{-0.08}$ & $          
0.71^{+0.16}_{-0.30}$ & $0.80^{+0.06}_{-0.07}$ & $31.04^{+33.77}_{-18.69}$ & $  
24.95^{+17.82}_{-14.10}$ & $5.70^{+8.65}_{-3.32}$ & $5.62^{+7.09}_{-2.96}$ \\ 
CO L400 & $[14-15]$ & $14.18$ & $12.06$ & $0.63^{+0.08}_{-0.07}$ & $          
0.74^{+0.14}_{-0.21}$ & $0.77^{+0.06}_{-0.07}$ & $24.80^{+31.20}_{-14.99}$ & $  
20.46^{+18.07}_{-11.60}$ & $5.61^{+6.63}_{-3.08}$ & $5.66^{+6.20}_{-2.95}$ \\ 
\hline
\end{tabular}

\label{tbl:sims_rhalf} 
\end{center}
\footnotetext{* $\newl (M/[\hMsun])$}
\vspace{-0.3in}
\end{minipage}
\end{table*}

Armed with the simulations described in \S~\ref{Sec:simulations} and with the technical definitions introduced in \S2.2-2.4, we present here a systematic study of the shape parameters. Specifically, we will present the shape parameters of the entire matter distribution in haloes (in \S\ref{Sec:ShapeHaloes}), and of different halo components (stars in \S\ref{Sec:ShapeStars} and hot gas in \S\ref{Sec:ShapeGas}) as well as their mass, spatial, and redshift dependence.
In Table \ref{tbl:sims_rhalf} we summarize the values and the scatter of the shape parameters and the misalignment angles for the stars within $r_{\rm half}^{\rm star}$.

It is well known that the reliability of shape estimates of particle distributions depends on the number of particles used to trace those distributions (e.g. Ten14). Motivated by the results presented in Appendix \ref{Sec:ShapeSampling}, 
we measure shapes of structures with at least 300 particles. { The resolution criterion is applied separately to the different halo components. Therefore, for a reliable shape measurement of the stellar component we require galaxies containing at least 300 stellar particles.} Our tests performed using synthetic NFW haloes show that this choice ensures a precision of 3\% and an accuracy better than 10\% in the sphericity and triaxiality parameters, see Appendix \ref{Sec:ShapeSampling} for more details. We note that Ten14 performed a similar convergence test according to which using 300 particles leads to $\sim -10 \%$ bias in the sphericity of a particle distribution. Our choice ensures relatively high precision while still allowing us to have a large number of haloes for which shape measurements can be performed.

\subsection{The shape of haloes}

\begin{figure*} \begin{center} \begin{tabular}{cc}
\includegraphics[width=1.0\columnwidth]{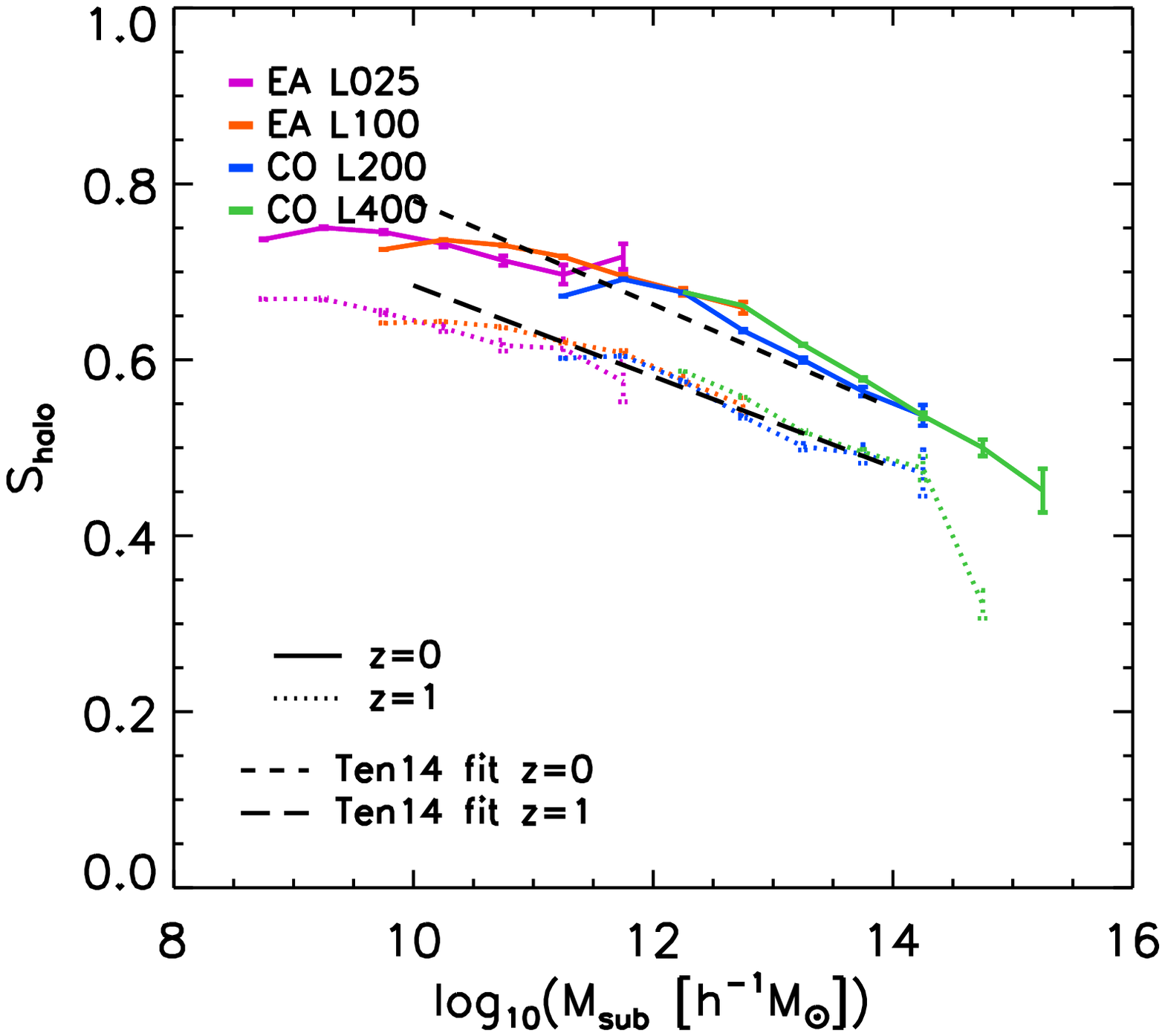} &
{\includegraphics[width=1.0\columnwidth]{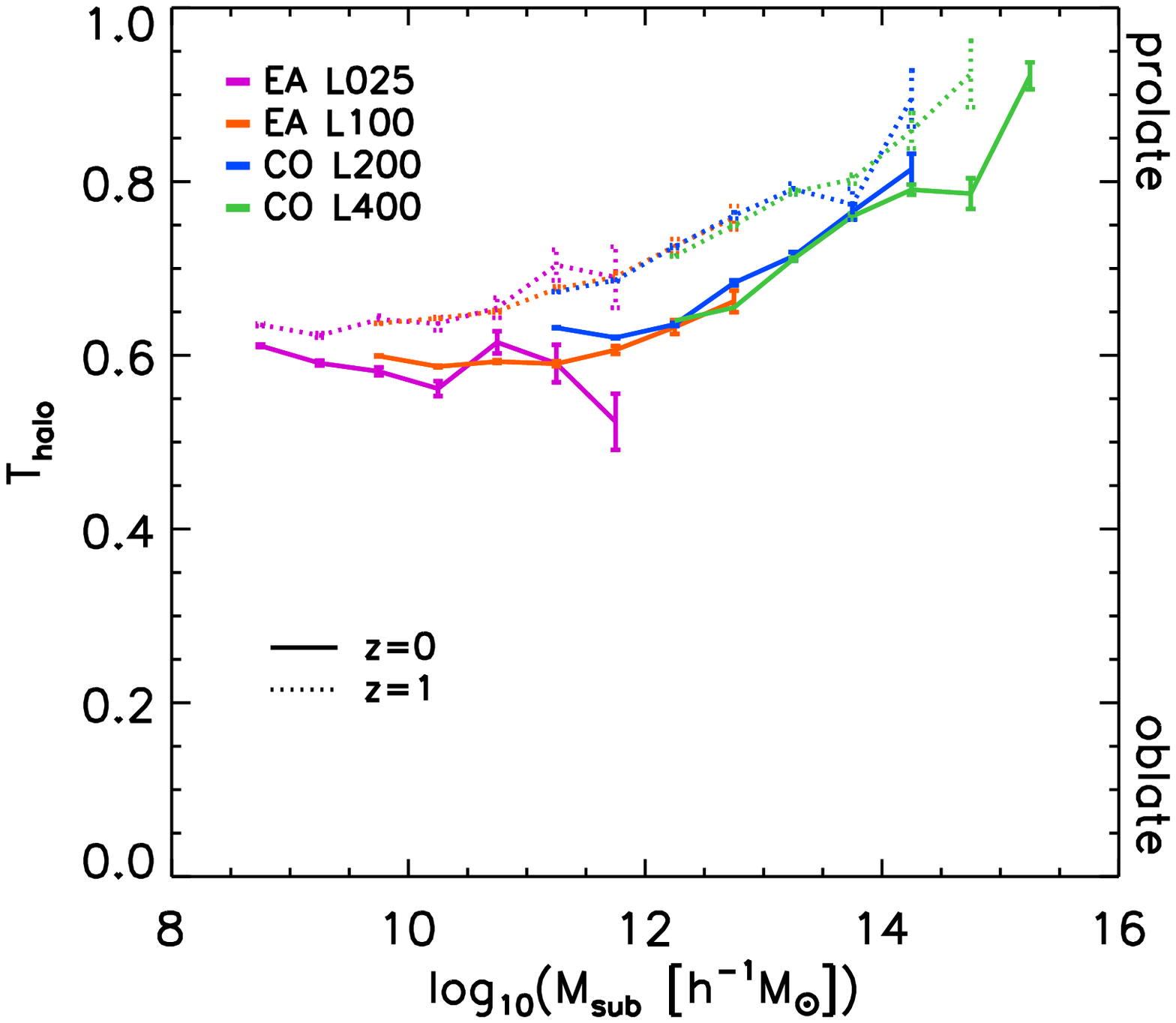}} \\
\end{tabular} \end{center}
\caption{ 
Halo sphericity (left) and triaxiality (right) as a function of halo mass.
Both central and satellite haloes are considered, hence the choice of $M_{\rm sub}$ (the sum of the masses all the particles belonging to the subhalo) as identifier of the halo mass. Both shape parameters are computed using all the particles in the subhaloes (gas, stars, and dark matter). Using only dark matter would give virtually identical results. Different colours indicate different simulations, whereas solid (dotted) lines refer to $z=0$ ($z=1$). The error bars represent the one sigma bootstrap error on the median. Dashed black lines are the values obtained using the fitting functions from Ten14.
}
\label{fig:ST_halo_vs_Msub}
\end{figure*} 

\label{Sec:ShapeHaloes}
Fig.~\ref{fig:ST_halo_vs_Msub} displays the sphericity (left panel) and triaxiality (right panel), $S$ and $T$ respectively, for halo masses in the range 
$9 \le \newl(M_{\rm sub}/[\hMpc])\le 15$. 
Different colours indicate different simulations and different line styles represent different redshifts (see legend). Notably, despite their difference in resolution, the results agree  in the overlapping mass intervals probed via different simulations. The common qualitative result is very simple: haloes become less spherical and more triaxial (prolate) with increasing mass. Sphericity (triaxiality) decreases (increases) from $z=0$ (solid lines) to $z=1$ (dotted lines). Haloes thus become more spherical/oblate as cosmic time progresses.
This effect is not due to baryon physics since it was also found in dark matter only simulations \citep[e.g.][]{Bryan12,Tenneti14a}. 
For comparison, we also plot the halo sphericity reported by \cite{Tenneti14a} using a dashed line for $z=0$ and a long dashed line for $z=1$. Despite the differences in box size, resolution and implementation of baryon physics, the overall agreement with our composite set of simulations is good at both redshifts.
The shape of the haloes when all particles are considered is dominated by the dark matter component. In fact, the shape of the dark matter component is nearly identical to the that of the total mass distribution. 

Our composite sample suggests that, over a wider range in halo masses, the relation deviates from linear showing a steepening from low to high masses.

\subsection{Shape of the stellar component of haloes}
\label{Sec:ShapeStars}

\begin{figure*} \begin{center} \begin{tabular}{cc}
\includegraphics[width=1.0\columnwidth]{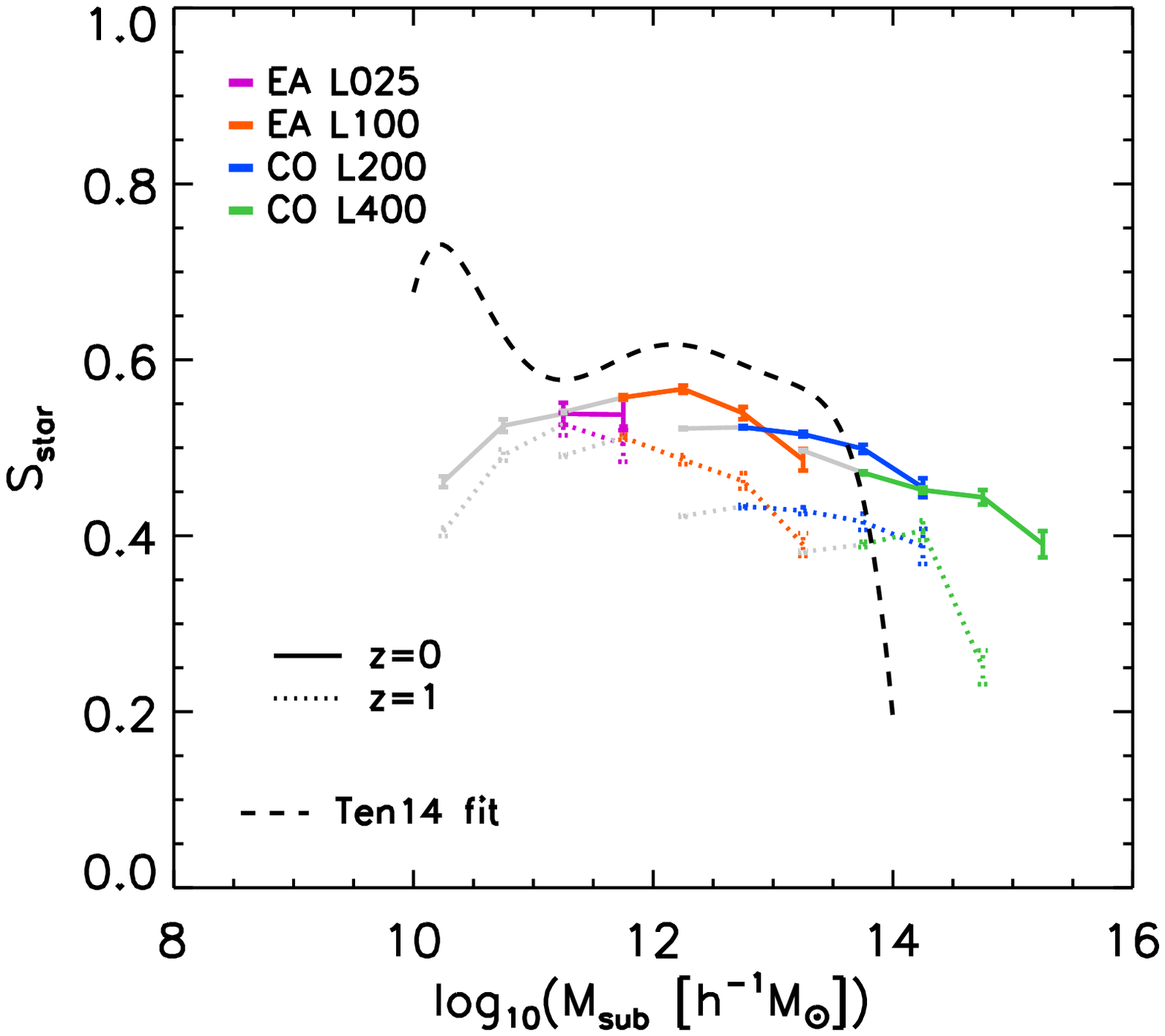} &
{\includegraphics[width=1.0\columnwidth]{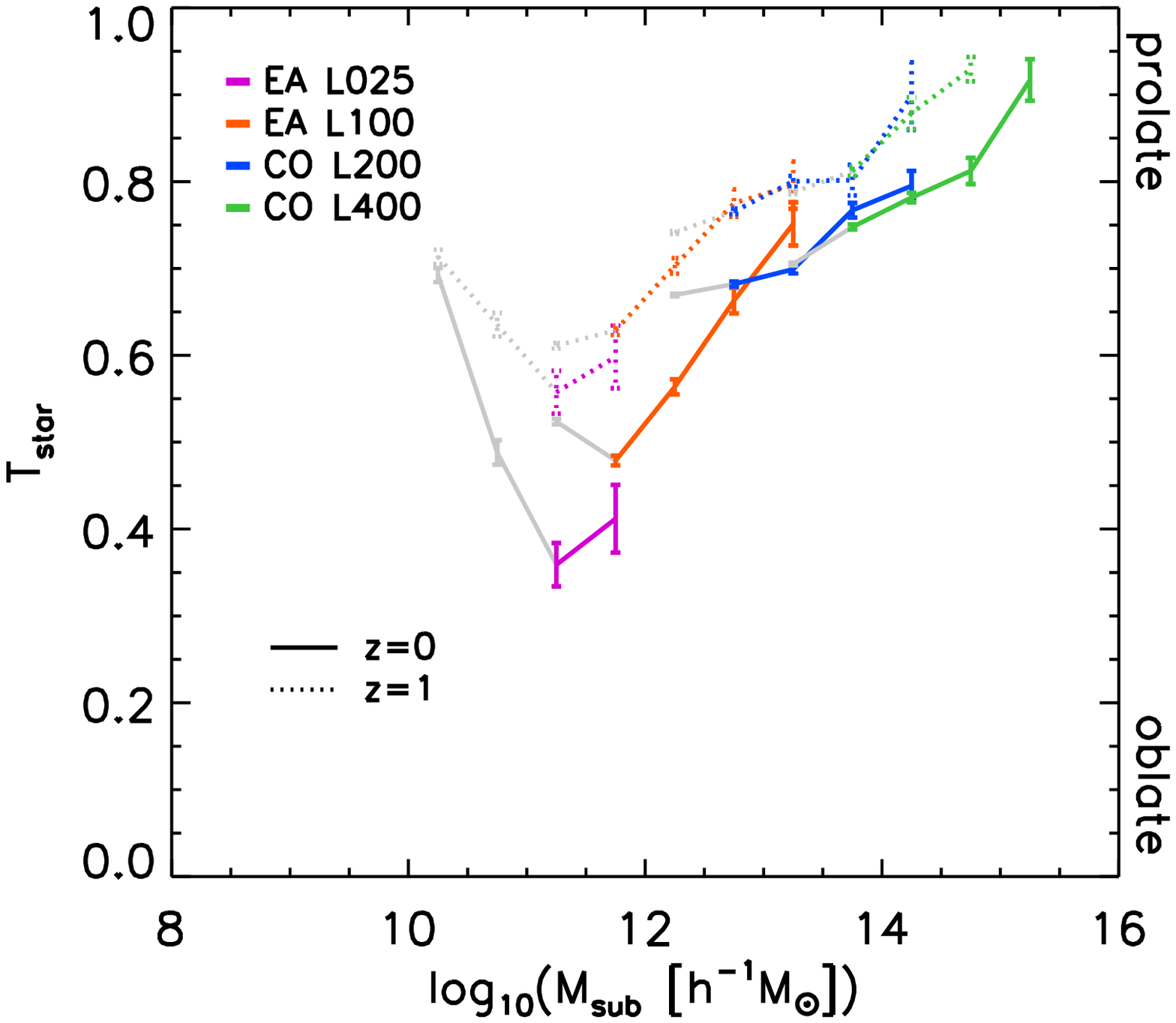}} \\
\end{tabular} \end{center}
\caption{ 
Stellar shape parameters (sphericity on the left, triaxiality on the right) 
as a function of halo mass.
Different colours indicate different simulations, whereas different line styles refer to different redshifts. The error bars represent the one sigma bootstrap error on the median. Grey lines are show the results for mass bins containing haloes with less than 300 stellar particles. The dashed black line indicates the sphericity obtained from the fitting function of Ten14. The upturn and the downturn in this fitting function are likely due to selection effects (see discussion in \S~\ref{Sec:star_STinR}).
}
\label{fig:ST_star_vs_Msub}
\end{figure*} 

Fig.~\ref{fig:ST_star_vs_Msub} displays the halo mass dependence of the shape parameters of the stellar distributions. Sphericity is on the left, triaxiality is on the right. 
As in Fig.~\ref{fig:ST_halo_vs_Msub} different colours indicate different simulations according to Table \ref{tbl:sims}.
We remind the reader that we use a minimum of 300 particles to determine the shape of particle distributions. This inevitably leads to a relatively small halo mass range for each simulation. 
However, the composite sample of our simulations covers the halo mass range $11 \le \newl(M_{\rm sub}/ [\hMsun]) \le 15$. Note that we have indicated with grey lines the values of the shape parameters obtained when considering haloes comprising fewer than 300 particles. Interestingly, in the overlapping halo mass range, the sphericity parameters derived from simulations with different resolutions agree remarkably well. 
The general trend seems to suggest that sphericity is a decreasing function of halo mass for $\newl(M_{\rm sub}/ [\hMsun]) > 12$ at $z=0$ and for $\newl(M_{\rm sub}/ [\hMsun]) > 11$ at $z=1$.

We compare our results in Fig.~\ref{fig:ST_star_vs_Msub} with the recent work of \cite{Tenneti14a} by showing their fitting function to the sphericity of the stellar component of haloes (black dashed line). 
The most prominent feature of their fitting function, namely the sharp upturn at masses $\newl(M_{\rm sub}/ [\hMsun]) < 11$, is most likely due to a selection bias. 
In their work they only compute shapes for subhaloes with more than 1000 stellar particles. 
This choice imposes a strict limit in stellar mass but not in subhalo mass. 
This approach only results in an unbiased selection if the minimum stellar mass of all haloes in a given mass bin is higher than $ > 1000 m_{\rm star}$ where $m_{\rm star}$ is the mass of a stellar particle. If we impose the same strict limit of 1000  star particles without also limiting the halo masses accordingly, we obtain a similar upturn in the stellar sphericity. Moreover, this upturn occurs at a different mass for different simulations since a fixed number of particles translates into different mass depending on the resolution used.

The triaxiality parameter (right panel of Fig.~\ref{fig:ST_star_vs_Msub}) is an increasing function of halo mass at both $z=0$ and $1$. As discussed in Appendix \ref{Sec:ShapeSampling}, the accuracy of the triaxiality estimate is more sensitive to the minimum number of particles used to compute it. This manifests itself in the fact that the grey lines in this plot do not continue a monotonic trend beyond the well-resolved mass interval, thus reinforcing the importance of imposing a minimum number of particles used when attempting to recover the triaxiality of a distribution of particles.

\subsubsection{The projected stellar rms ellipticity}
\label{Sec:erms}

\begin{figure} \begin{center} 
\includegraphics[width=1.0\columnwidth]{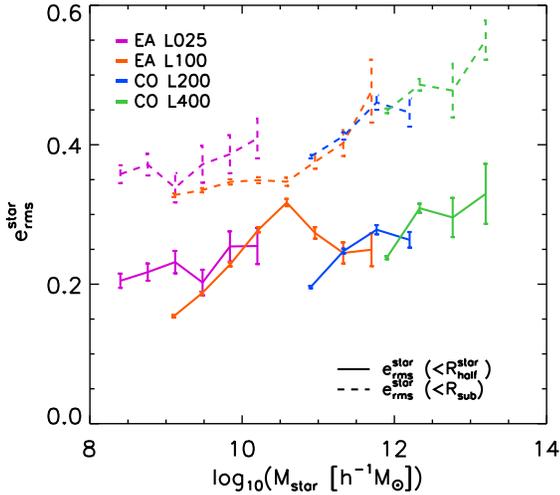} 
 \end{center}
\caption{
Projected rms stellar ellipticity as a function of halo mass. 
Different colours indicate different simulations (and therefore halo masses), whereas different line styles indicate the region within which the stellar distribution is considered. Specifically, dashed lines indicate the case in which only star particles within the entire halo are considered, whereas solid lines indicate the case in which star particles within the stellar half-mass radius are considered. Both centrals and satellites are considered this analysis.
}
\label{fig:erms_vs_Msub}
\end{figure}

Under the assumption that galaxies are randomly oriented, averaging the observed projected ellipticity of galaxies gives a measurement of the gravitational lensing effect that, in turn, gives constraints on the matter distribution along the line of sight. The S/N of those measurements depend on the second moment of the distribution of galaxy intrinsic ellipticity, termed $e_{\mathrm{rms}}$. Many observational studies have measured the value of the \erms for populations of galaxies. Early results were reported in \citealt[][]{Hoekstra00}, and more statistically robust results were obtained using SDSS data \citep[][]{Reyes12}, COSMOS \citep[][]{Joachimi13, Mandelbaum13} and the CFHTLenS survey \citep[][]{Heymans13,Miller13}.
Unfortunately, despite the tremendous progress in the statistical power of the galaxy surveys employed in these studies, obtaining an accurate estimate of \erms remains challenging, especially because of the fact that the quantity that is accessible observationally always has a (often non-negligible) noise contribution \citep[see e.g.][]{Viola14}.
 
For our composite sample of haloes, \erms is defined as:
\begin{equation} \label{eq:meanrms}
 e^2_{\rm{rms}} = \frac{1}{N}  \sum_{i}{ \left( \frac{1 - q_{i}'^2}{1 + q_{i}'^2} \right)^2 }
\end{equation}
where $q_{i}'$ is the projected ellipticity of the $i^{th}$ halo $q' = {b'}/{a'}$ where $a'$ and $b'$ are the values of the major and minor axis of the projected stellar distribution, and $N$ is the total number of haloes considered. 

We use our composite sample of haloes to compute the stellar \erms in bins of halo mass of width $0.5 ~ \rm dex$, as a function of halo mass in Fig~\ref{fig:erms_vs_Msub}.
We make use of all star particles that belong to the subhaloes (dashed lines) or only stellar particles within the stellar half-mass radius (solid lines). Both centrals and satellites are considered for this analysis.
{ When all star particles are considered the value of the \erms increases with mass from 0.35 to 0.55. Systematically lower values are found if only stellar particles within the half-mass radius are considered.}

{ The values of the \erms predicted by our composite sample, when all stars are considered, are in broad agreement with the observed noise-corrected values that are of the order of $\approx 0.5$-$ 0.6$ depending on luminosity and galaxy type \citep[e.g.][]{Joachimi13}.} Unfortunately, a direct comparison of our results with those obtained from observational studies is far from trivial. In fact, it would be crucial to mimic all steps in the observational methodology. For instance, \erms measurements are usually only available for a given sub-population of galaxies, those galaxies are further binned in absolute magnitude, and the axis ratio is computed starting from (noisy) images for which flux isophotes need to be identified. In the context of this investigation, we find the current level of agreement satisfactory and ideal as a starting point for future explorations.

\subsubsection{Variation of the shape of the stellar component of haloes with the distance from the halo centre}
\label{Sec:star_STinR}

\begin{figure*} \begin{center} \begin{tabular}{cc}
\includegraphics[width=1.0\columnwidth]{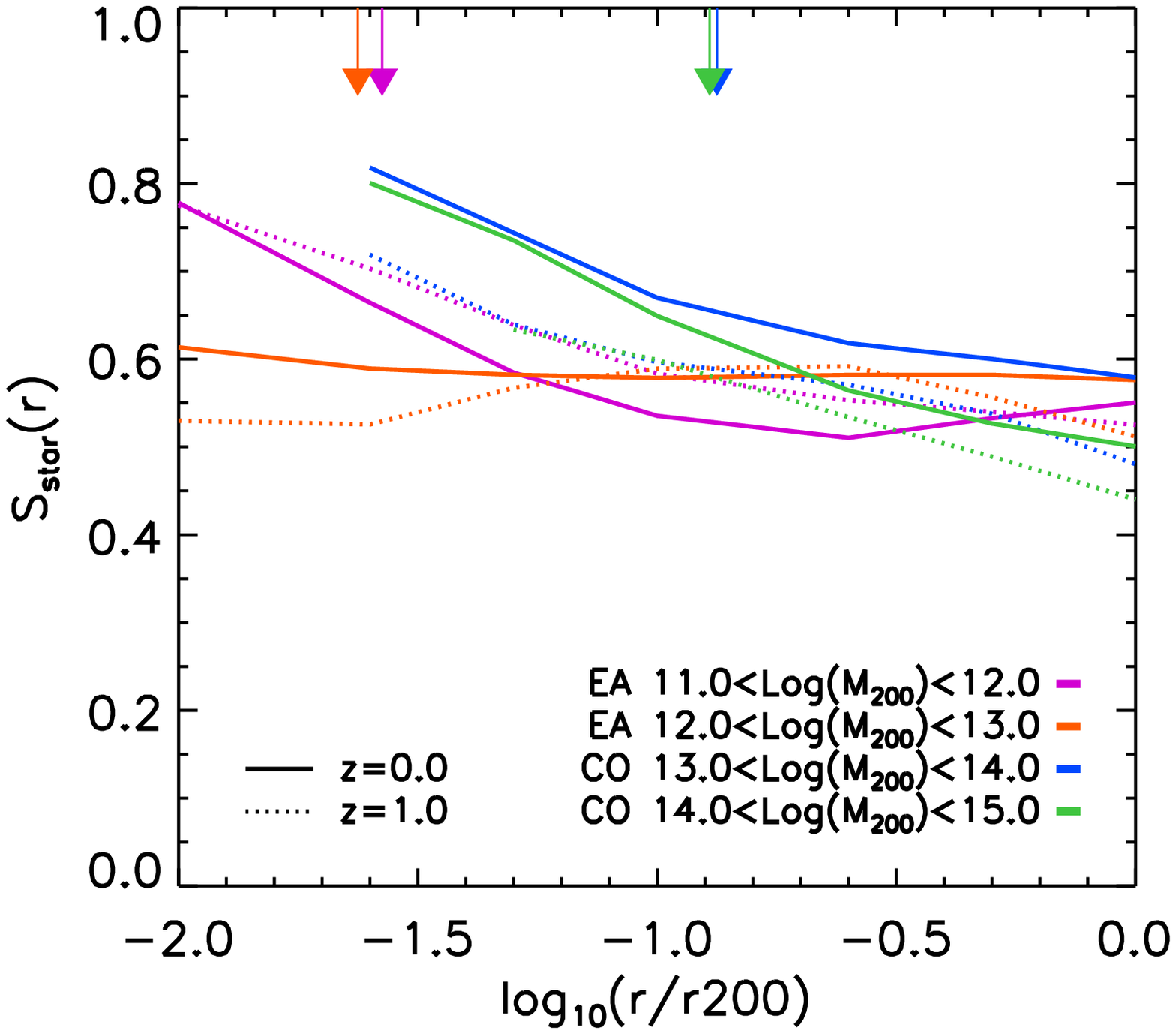} &
\includegraphics[width=1.0\columnwidth]{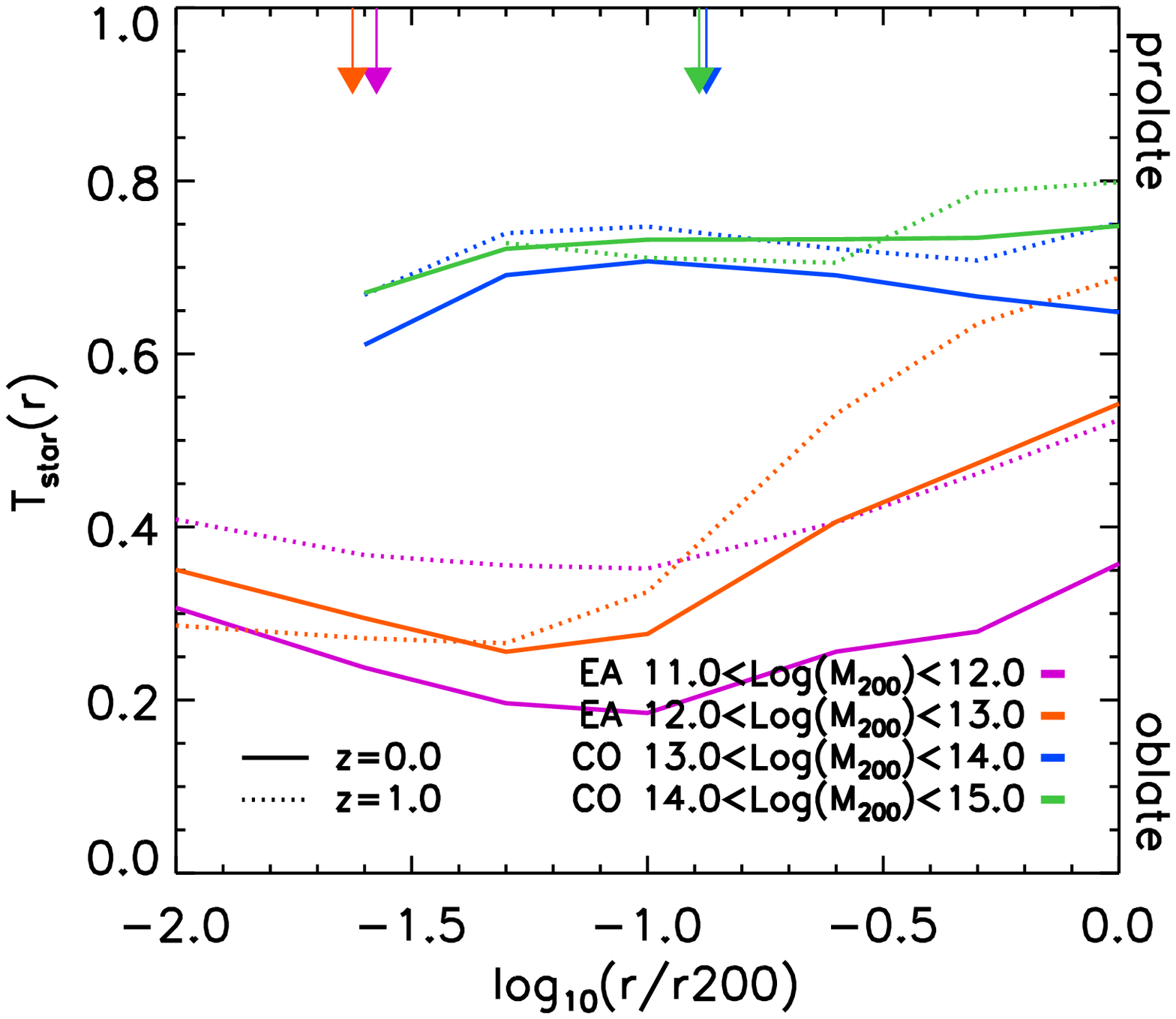} \\
\end{tabular} \end{center}
\caption{
Stellar shape parameters (sphericity on the left, triaxiality on the right) as a function of distance from the centre of the halo for haloes in different halo mass bins (see legend). Distances have been rescaled to the mean halo radius in each mass bin, $r_{200}^{\rm crit}$ to ease the comparison of the results for different masses. Different colours indicate different simulations, whereas different line styles refer to different redshifts.
The distribution becomes less spherical and more prolate with increasing distance from the halo centre.
Vertical arrows indicate the median values of the half-mass radii in stars, $r_{\rm half}^{\rm star}$, which can be considered a proxy for the typical extent of a galaxy. The blue arrow
lies beneath the green one.
}
\label{fig:star_STinR}
\end{figure*} 
Fig.~\ref{fig:star_STinR} shows the sphericity (left panel) and the triaxiality (right panel) of the stellar component of haloes as a function of the distance from the centre of the halo. 

We divide our sample into mass bins that are drawn from different simulations according to Table \ref{tbl:sims}.
We then compute the inertia tensor for increasingly larger spheres around the centre of each halo. 
For every sphere we show the median values of the shape parameters of the mass inside the sphere.
Radii are given in units of $r_{200}^{\rm crit}$ to allow for a comparison of haloes of different masses.
Only particles that are bound to the halo are considered for this analysis. Curves are drawn only on scales where at least 300 particles can be used.

The stellar component of haloes tends to be more spherical near the centre. The triaxiality value shows significant evolution for masses below $M_{\rm sub}<10^{12} \hMsun$.
These trends are qualitatively the same as those found for the dark matter component (not shown) with the exception that the radial profile of the stellar distribution is steeper than that of the dark matter distribution.

The right panel of Fig.~\ref{fig:star_STinR} shows a large difference between the triaxiality values of subhaloes in the mass bins $12<\newl(M_{\rm sub}/ [\hMsun]) < 13$ (orange curves, EA L100) and $13<\newl(M_{\rm sub}/ [\hMsun])<14$ (blue curves, CO L200). This feature might be caused by the different resolution, volume, and/or baryon physics of the two sets of simulations (although the latter is relatively small). To test whether that is the case, we  compute the triaxiality parameter of subhaloes with mass $13<\newl(M_{\rm sub}/ [\hMsun])<14$ using the EAGLE L100 simulation (not shown). We find the corresponding results to agree with the results obtained using  the same mass bin from the cosmo-OWLS L200 simulation. Thus, we interpret  the differences between the triaxiality of subhaloes in the mass bins $12<\newl(M_{\rm sub}/ [\hMsun])< 13$ and $13<\newl(M_{\rm sub}/ [\hMsun])<14$ as having a physical origin rather than being due to the resolution, the volume, or the (small) differences in the baryon physics of the two sets of simulations.

\subsection{Shape of the hot gas component of haloes}
\label{Sec:ShapeGas}

\begin{figure*} \begin{center} \begin{tabular}{cc}
\includegraphics[width=1.0\columnwidth]{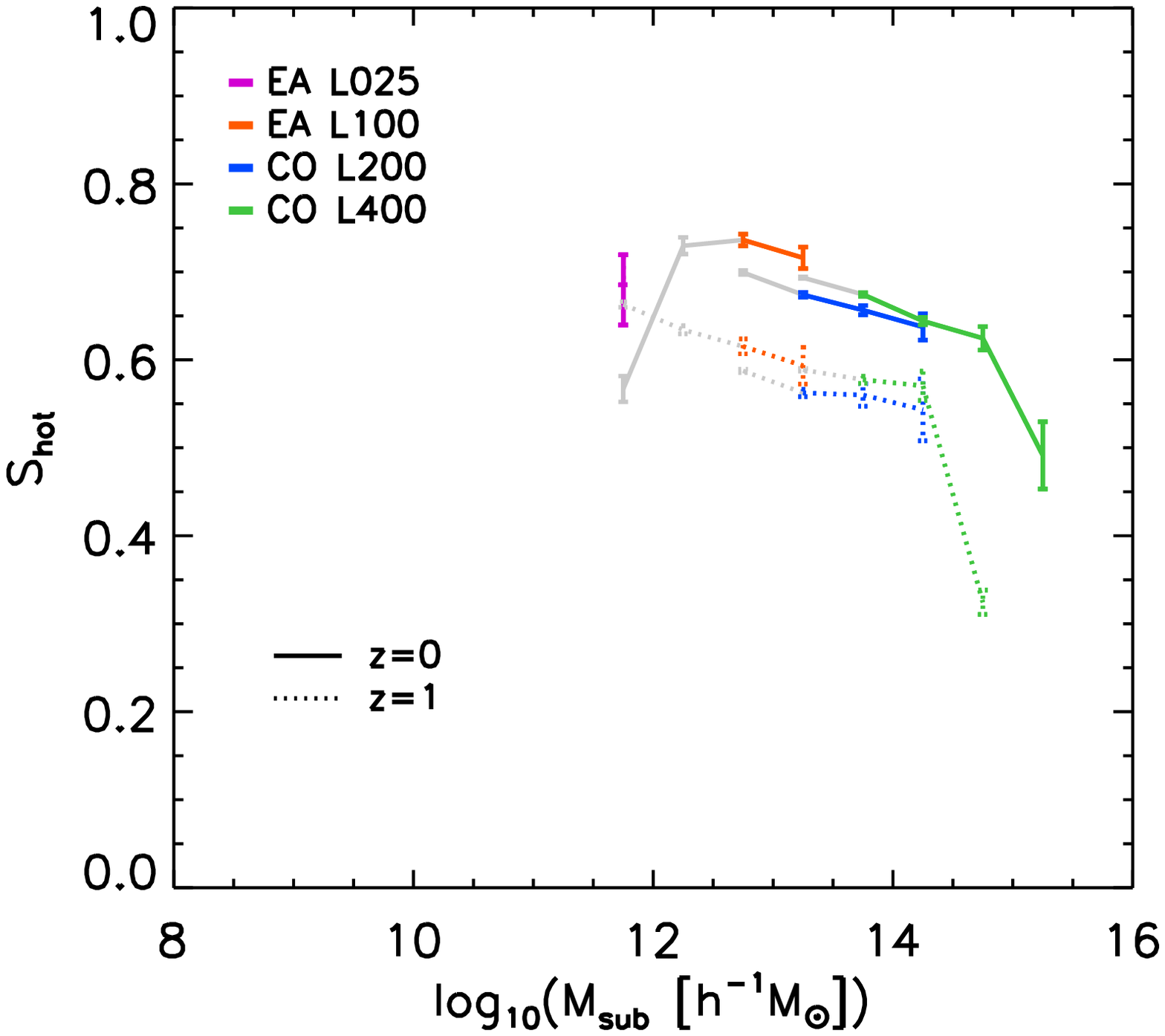} &
{\includegraphics[width=1.0\columnwidth]{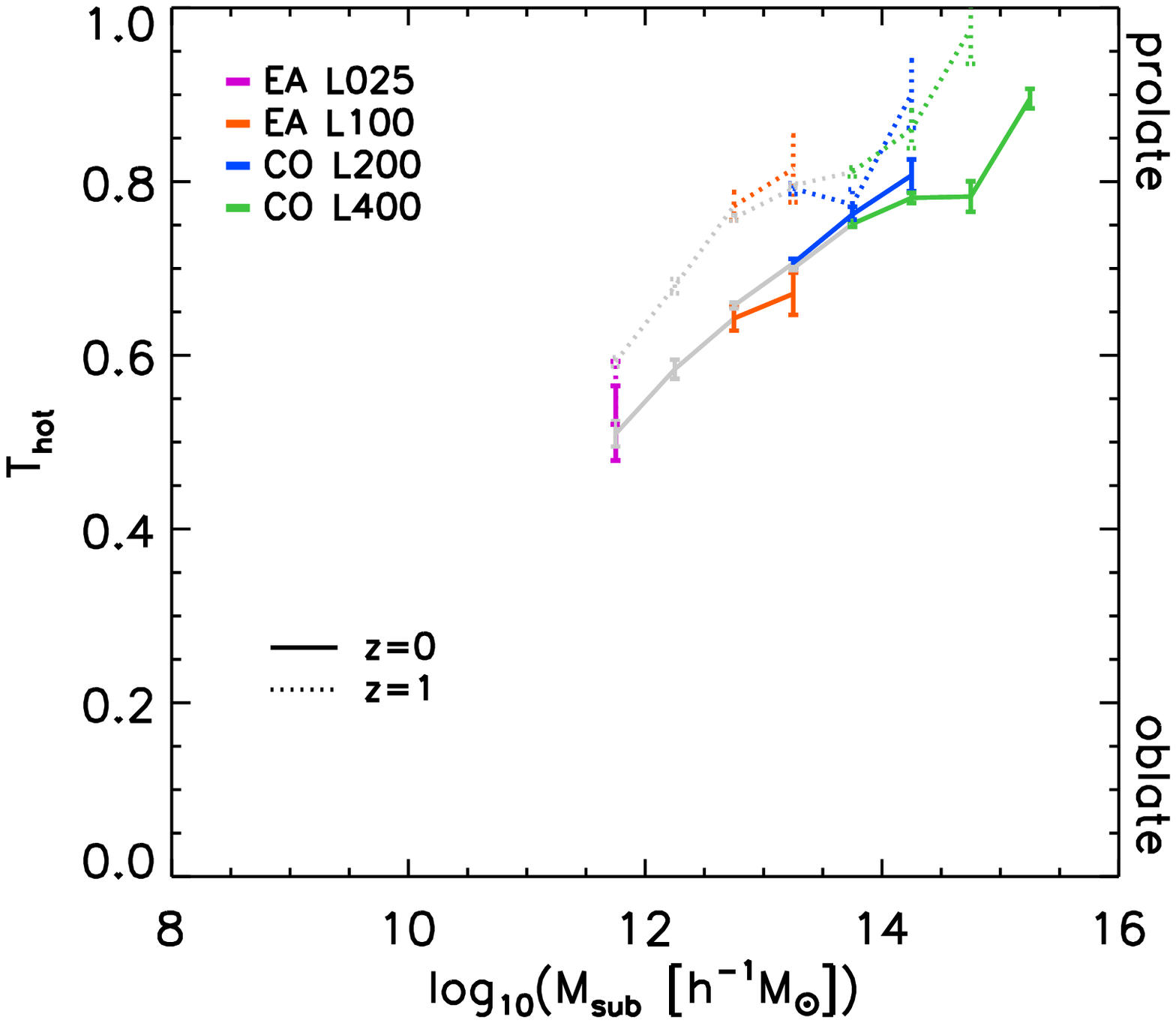}} \\
\end{tabular} \end{center}
\caption{
Shape parameters (sphericity on the left, triaxiality on the right) of the gas distribution  for the hot ($T>10^6 \rm K$) component.
Different colours indicate different simulations, whereas different line styles refer to different redshifts. The error bars represent one sigma bootstrap error on the median. Grey lines show the results for mass bins containing haloes with less than 300 hot gas particles. 
}
\label{fig:ST_hot_vs_Msub}
\end{figure*} 

In this section we repeat, for the hot gaseous component of haloes, the analysis performed for the total and stellar matter in \S~\ref{Sec:ShapeHaloes} and \S~\ref{Sec:ShapeStars}, respectively.
We present the shape parameters for a subsample of temperature-selected diffuse gas ($T>10^6 \rm K$). The selection is quite insensitive on the exact temperature cut, since most of the hot gas in groups and cluster has a temperature that is a factor of two greater than the virial temperature.
This temperature selection is used as a rough proxy for the hot X-ray emitting gas. A proper selection of X-ray emitting gas is beyond the scope of this paper, as this would require an accurate computation of the X-ray luminosity of the gas particles. 
A luminosity weighted scheme for the shape of the hot gas would result in the inner regions dominating the shape resulting in more spherical shapes \citep[][]{Crain13}.  On the other hand, a mass weight scheme, as adopted in this work, would be closer to the shape that a  Sunyaev-Zeldovich (SZ) experiment would measure since SZ flux is proportional to the gas mass and the temperature, making it potentially testable with a combined SZ-lensing analyses.

Fig.~\ref{fig:ST_hot_vs_Msub} presents the variation of the shape parameters, sphericity on the left and triaxiality on the right, of the temperature-selected hot gas particle. 
The convergence of the sphericity parameter between the different simulations is poorer in this case than for other components shown earlier. By imposing a strict limit on the number of particles needed for measuring the shape we limit our results to only few points for the EAGLE simulations. For instance, is no longer possible to connect the results from L025 and L100. Nonetheless by relaxing the constraint on the number of particles (grey points), it is possible to identify a trend in the shapes that suggests an increasing triaxiality and decreasing sphericity of the hot gas component with host halo mass. 

We have also studied the radial dependence of the shape parameters for the hot gas component of haloes (not shown). Given the limit on the minimum number of particles, only three mass bins could be  investigated ($M_{200}^{\rm crit}> 10^{12} \hMsun$) and only down to radius of $r/r_{200}^{\rm crit} = 0.3$, for which no significant radial trend was found.

\section{Misalignment of Galaxies with their own host haloes}
\label{Sec:MisAlignment}

In this section, we show the relation between the orientation of haloes and that of their stellar and hot gas component. 
Specifically, we will show how the orientation of the major axis of the stellar distribution (\S\ref{sec:starmisalignment}) and of the hot gas distribution (\S\ref{sec:gasmisalignment}) compare to that of the host halo. Similarly to the case of the shape parameters, we will investigate the mass, radial, and redshift dependence of this relation. 
Note that we focus our study mainly on central haloes. 
We remind the reader that a formal definition of the axes of particle distribution and their relative misalignment angles is provided in \S\ref{Sec:MisAlignmentDefinition}.

\subsection{Misalignment of stars with their host haloes}
\label{sec:starmisalignment}

\begin{figure*} \begin{center} \begin{tabular}{cc}
\includegraphics[width=1.0\columnwidth]{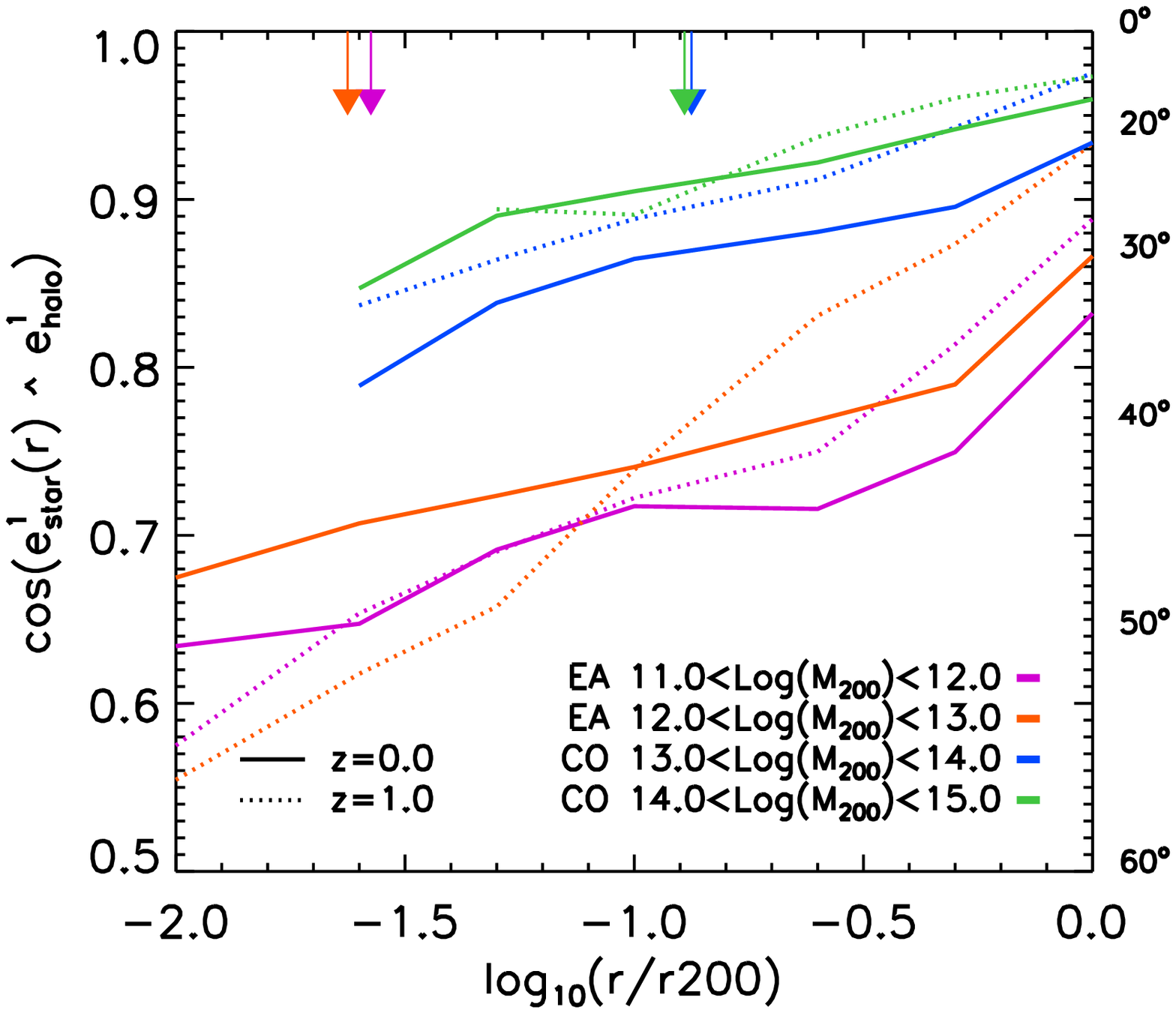} &
{\includegraphics[width=1.0\columnwidth]{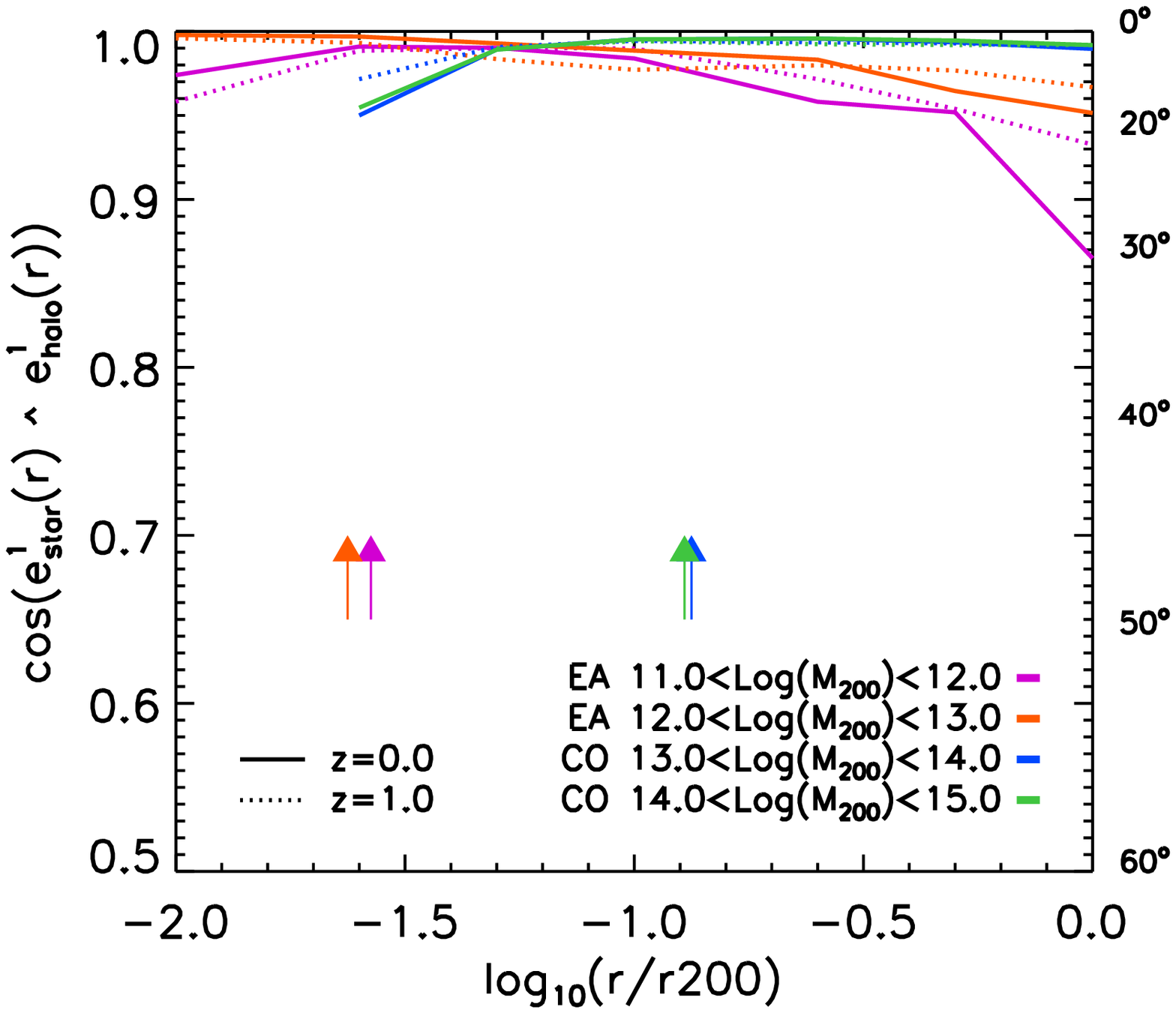}} \\
\end{tabular} \end{center}
\caption{
Spatial variation of the median cosine of the misalignment angle between the major axis of stars and the underlying (mostly dark) matter distribution. Different colours indicate different halo mass bins, whereas different line styles indicate different redshifts. Radial coordinates are normalized by the mean halo radius, $r_{200}^{\rm crit}$, of each mass bin to ease the comparison. Only central haloes are used.
Vertical arrows represent the median value of $r_{\rm half}^{\rm star}$ in units of $r_{\rm crit}^{200}$ in  different mass bins.
{\it Left Panel.}
Median value of the cosine of the angle between the major axes of the stellar component and that of the entire halo. Here the direction of the halo is determined using all particles belonging to the halo. 
{\it Right Panel.}
Median value of the cosine of the angle between the major axes of the stellar component and that of the halo. The misalignment between the stars and halo is caused, to first order, by the misalignment of the inner dark matter halo with the total matter distribution in the halo. 
}
\label{fig:star_vs_halo_MinR_orient}
\end{figure*} 

The left panel of Fig.~\ref{fig:star_vs_halo_MinR_orient}
shows the median
misalignment of stars in spheres of increasingly larger radii with the direction of the 
total matter distribution within the virial radius for different bins in halo mass and for 
radii expressed in units of $r_{200}^{\rm crit}$. Perhaps not surprisingly, the alignment of
stars within the total halo increases from the inner to the outer part of the halo. The 
gradient is relatively steep, with the misalignment angle between the stars and their host 
haloes decreasing from about 30 degrees (at $r\sim 0.03 r_{200}^{\rm crit}$) to a few 
degrees (at $r\sim r_{200}^{\rm crit}$) in the case of the most massive haloes. In less 
massive haloes, the misalignment is larger at all scales. Similar trends hold at $z=1$ 
(dotted lines). The right panel of Fig.~\ref{fig:star_vs_halo_MinR_orient} shows the
misalignment of stars with the direction of the total (mostly dark) matter, where both are now 
enclosed in spheres of increasingly larger radii. At each radius, the misalignment is small. Stars are aligned with the total mass to within a few degrees in the most massive haloes,  
whereas the alignment deteriorates to about 20-30 (10-20) degrees for the least 
massive haloes at $z=0$ ($z=1$).

The misalignment of stars with their host halo can vary substantially depending on the
radius and the mass of a halo. The arrows in the plot represent the values, in units of 
$r_{200}^{\rm crit}$, of the half mass radius in stars, which is a good indicator of the 
physical extent of a galaxy. At this radius the orientation of the galaxies is clearly a 
biased proxy of the orientation of the halo. Galaxies are, however, much better aligned 
with the \emph{local} distribution of matter. 
This indicates that the the stellar orientations follow that of the dark matter, 
which is the dominant component in mass, and the dark matter itself changes 
orientation from the inner to the outer halo. This causes the stars to be well aligned 
with the local mass distribution but misaligned with the orientation of the entire halo.

\subsubsection{Probability distribution function of misalignment angles}
\begin{figure*} 
\begin{center} \begin{tabular}{c}
\includegraphics[width=2.0\columnwidth]{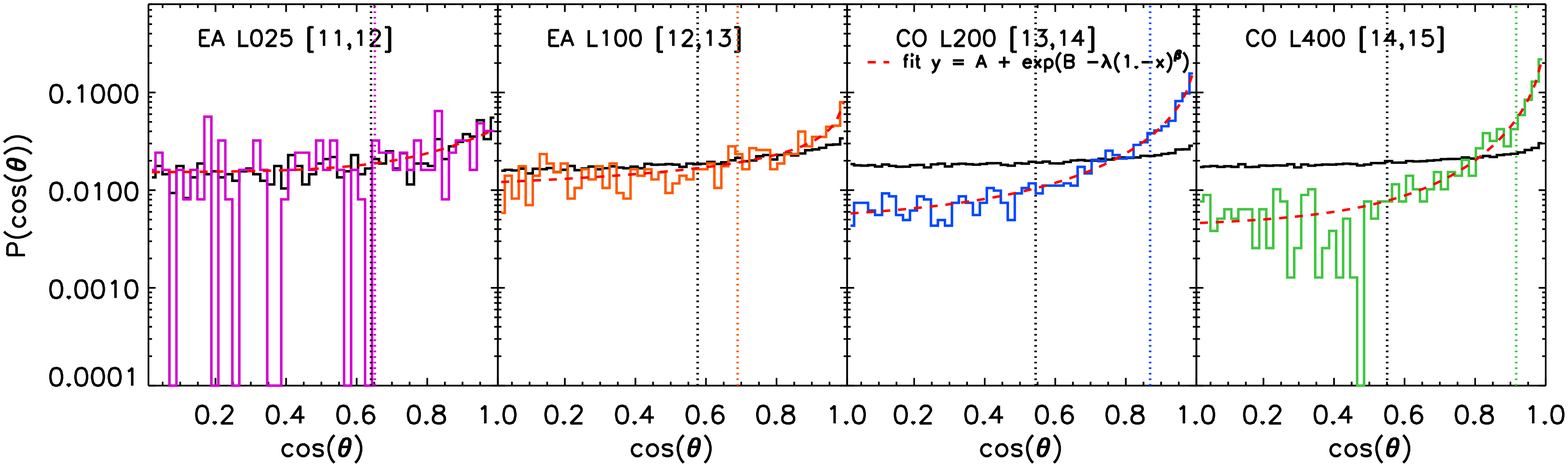}\\
\includegraphics[width=2.0\columnwidth]{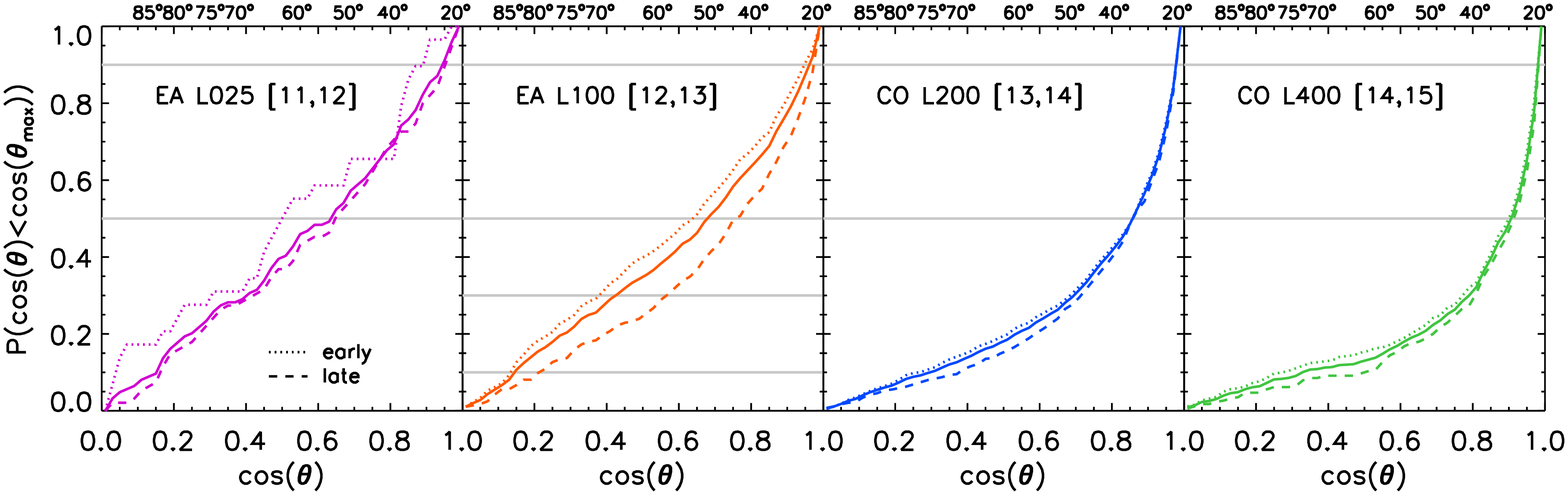}\\
\end{tabular}
\end{center}
\caption{
Upper panel: probability distribution function of the cosine of the misalignment angle between the major axis of the distribution of stars inside $r_{\rm half}^{\rm star}$, and the major axis of the entire halo for four halo mass bins.
The black histograms indicate the probability distributions for the total sample of haloes that satisfies the resolution criteria, whereas coloured histograms refer only to the subsample of haloes whose mass is indicated in the legend.
Vertical lines indicate the median values of the misalignment angle (same colour convention as for the histograms). Red dashed curves represent the analytic fit discussed in Appendix~\ref{Sec:fits}.  
{ Lower panel: cumulative version of the probability function for early- and late-type galaxies (dotted and dashed curves respectively).}
}
\label{fig:star_inhalo_orient_histo}
\end{figure*} 

\begin{figure*} 
\begin{center} \begin{tabular}{c}
\includegraphics[width=2.0\columnwidth]{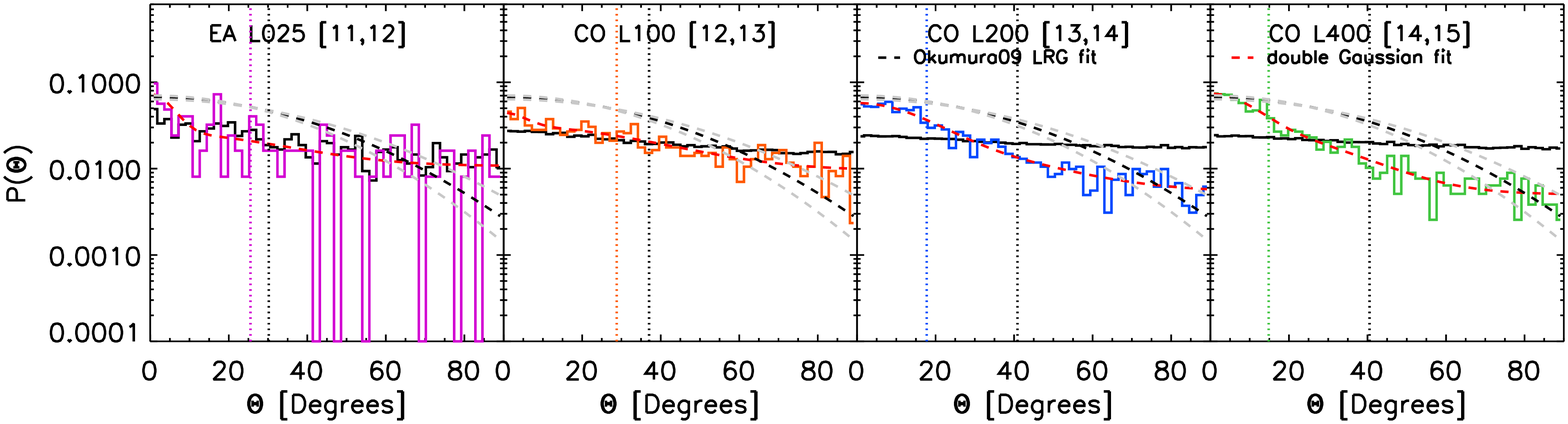}\\
\end{tabular}
\end{center}
\caption{
Probability distribution function of the 2D misalignment angle between the major axes of the {\it projected} distribution of stars (inside $r_{\rm half}^{\rm star}$) and the major axes of the {\it projected} total matter distribution for four halo mass bins. The black histograms indicate the probability distributions for the total sample of haloes that satisfies the resolution criteria, whereas coloured histograms refer only to the subsample of haloes whose mass is indicated in the legend.
Vertical lines indicate the median values of the misalignment angle (same colour convention as for the histograms).
Red dashed curves represent the analytic fit discussed in Appendix \ref{Sec:fits}, whereas black and grey curves are obtained with analytic functional forms that have been employed in the literature (see text).
}
\label{fig:star_inhalo_orient_histo_2d}
\end{figure*} 

In the previous section we presented the median value of misalignment between the halo and the stellar component.
The upper panel of Fig.~\ref{fig:star_inhalo_orient_histo} shows the probability distribution function of 
the cosine of the misalignment angle between the stars and the entire host halo for central 
galaxies. Here the stars are taken to be inside $r_{\rm half}^{\rm star}$. Each panel 
shows a different mass bin and therefore a different simulation. The colour histograms 
show the misalignment distribution for haloes in that specific mass bin, whereas the 
black histograms show the probability distribution functions for all haloes that are above the halo mass resolution limit (300 stellar 
particles inside $r_{\rm half}^{\rm star}$) in the 
corresponding simulation .
The vertical lines show the median values for the distributions and the dashed red curves
are analytic fits (see Appendix \ref{Sec:fits}).
{ The lower panel of Fig.~\ref{fig:star_inhalo_orient_histo} shows the cumulative probability of the cosine of the misalignment angle for early- (dotted curves) and late-type galaxies\footnote{See definition of disc galaxies in \S\ref{Sec:galtype_star_vs_halo_MinR_orient}.} (dashed curves) as well as for the whole sample of haloes (continuous curves).}

The distribution of the cosine of the misalignment angle has a long tail towards low 
values (i.e. strong misalignment) with a floor value that decreases with increasing halo
mass. The misalignment angle distribution of resolved haloes is quite similar in shape 
for the different simulations. Using the fitting
functions provided in Appendix \ref{Sec:fits} and the median values of the misalignment
angle shown in the previous plots, it is possible to populate dark matter haloes with galaxies oriented such that these misalignment distribution are reproduced.

\citet{Bett10} quantified the misalignment angle between the stellar and total matter distribution in a sample of (about 90) \emph{disc} galaxies selected from a hydrodynamic simulation in a cubic volume of $35\hMpc$ by side. They found that half of these galaxies have a misalignment angle larger than 45 degrees. Using the GIMIC simulations \citep[][]{gimic09}, \citet{Deason11} reported that 30\% of disc galaxies with average halo mass of $\newl(M_{\rm sub}/[\Msun])=12.1$
have a misalignment angle of more than 45 degrees.
Both these studies are in broad agreement with our findings for similar halo masses. Specifically, in the EAGLE simulations, we find that half of the disc galaxies have misalignment angles larger than 50 (40) degrees in L025 (L100) and 30\% of the galaxies in L100 (for which the typical halo mass is close to that in \citet{Deason11} have misalignment angles larger than 60 degrees.

Fig.~\ref{fig:star_inhalo_orient_histo_2d} shows the probability distribution 
function of the misalignment between the major axes of the \emph{projected} halo and the 
projected stellar mass component. For comparison, we report with black (mean) and grey (one sigma deviation)
dashed lines the results from \citet{Okumura09} who found that, by assuming a Gaussian
misalignment distribution between LRGs and dark matter haloes, they were able to 
account for the discrepancy between the measured orientation correlation of LRGs
and the one predicted by N-body simulations. Furthermore, we overplot analytic fits
to our discrete distributions using a double Gaussian (red dashed curves, see Appendix \ref{Sec:fits}).
Notably, none of our probability distributions resembles a Gaussian function. 
It is obvious that a single Gaussian function cannot be used as a fair description 
of the probability functions measured from our simulations. 

\subsubsection{Misalignment for early- and late-type galaxies}
\label{Sec:galtype_star_vs_halo_MinR_orient}

\begin{figure*} \begin{center} \begin{tabular}{cc}
\includegraphics[width=1.0\columnwidth]{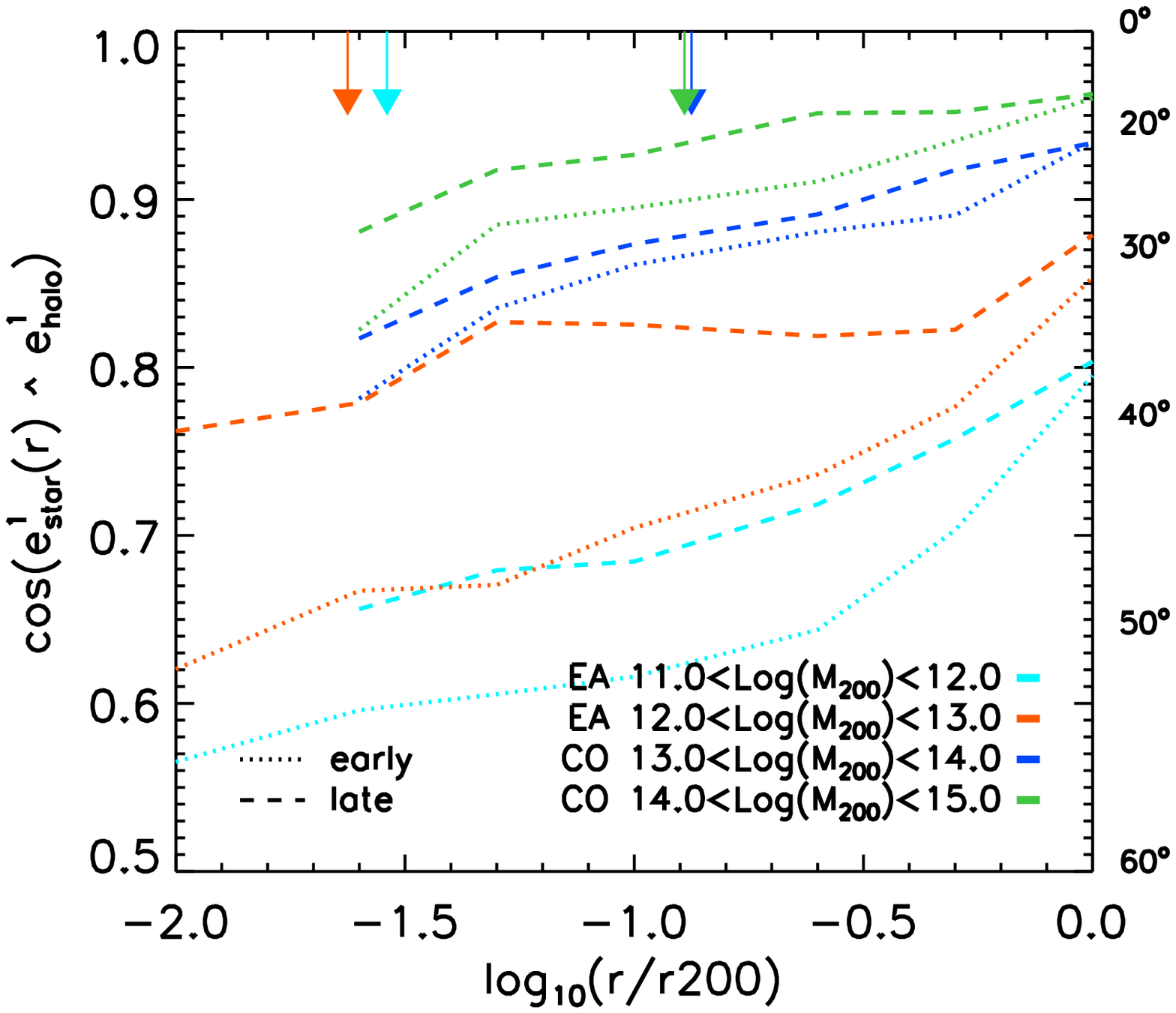} &
{\includegraphics[width=1.0\columnwidth]{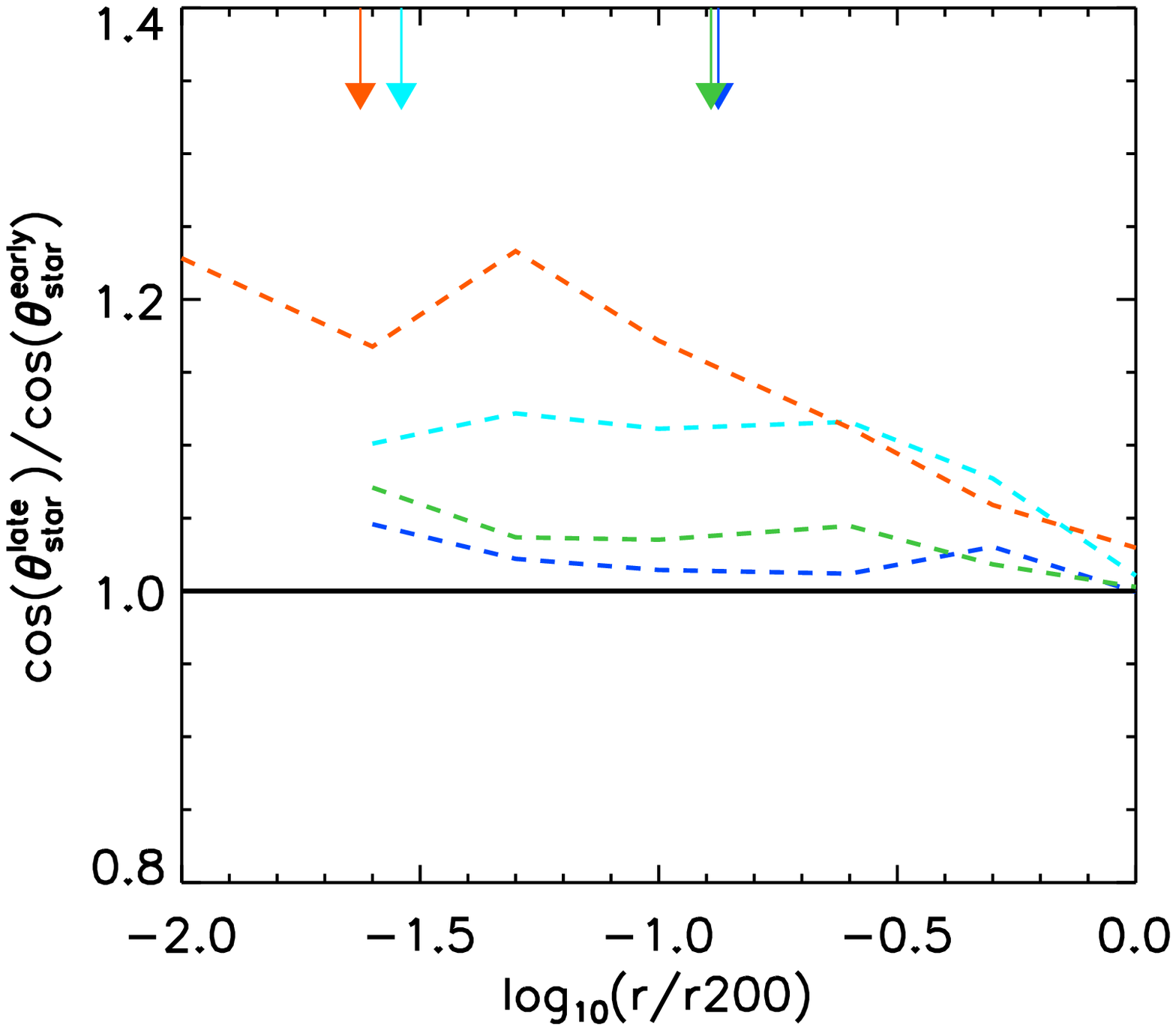}} \\
\end{tabular} \end{center}
\caption{
{\it Left Panel.}
Spatial variation of the cosine of the median misalignment angle between stars and the underlying (mostly dark) matter distribution. Different colours refer to different halo mass bins, whereas different line styles refer to different galaxy types. Radii are rescaled to the mean halo radius, $r_{200}^{\rm crit}$, to ease the comparison of the results corresponding to different halo mass bins. Only central haloes are used in order to remove effects that can alter mostly the alignment of satellites. Here the direction of the halo is determined using all particles belonging to the structure. A kinematic classification has been employed (see text) to divide galaxies into early- and late-type. 
{\it Right Panel.} Spatial variation of the ratio between the alignment 
of late and early type galaxies. As in the left panel, the alignment is expressed in terms of the cosine of the angle between galaxy and halo major axes.   
In both panels the least massive bin is taken from the L100 simulation to improve on the otherwise poor statistics of L025.
The vertical arrows represent the median values of $r_{\rm half}^{\rm star}$ in units of $r_{\rm crit}^{200}$ in different mass bins.
}
\label{fig:galtype_star_vs_halo_MinR_orient}
\end{figure*} 

In this section, we study the alignment between stars and their host haloes in early- and 
late-type galaxies. Given the galaxy stellar velocity dispersion,
$\sigma_{\rm star}$, and the halo maximum circular velocity, $V_{\rm circ}^{\rm max}$, 
one can define the ratio $\eta = \sigma_{\rm star}/V_{\rm circ}^{\rm max}$ to 
quantify whether a galaxy is supported either by ordered (rotational) motion 
or by the velocity dispersion. We adopt the convention that $\eta \leq 0.5$ 
indicates a rotationally-supported galaxy (late type), whereas  $\eta > 0.5$ indicates a dispersion-supported galaxy (early type). 

The left panel of Fig.~\ref{fig:galtype_star_vs_halo_MinR_orient} shows the median misalignment of the direction of the entire host halo with that of stars in spheres of increasingly larger radii for  early- (dotted lines) and late-type (dashed lines) galaxies. As for the entire galaxy population, the misalignment of stars with their host halo decreases from the inner to the outer part of the halo. The misalignment decreases with mass and is lower for late- than for early-type galaxies. The misalignment of early-type galaxies in low-mass haloes\footnote{In Fig.~\ref{fig:galtype_star_vs_halo_MinR_orient}, we use the EAGLE L100 simulation also for the least massive bin (cyan lines) to improve the otherwise poor statistics of the EAGLE L025 simulation.} ($11<\newl (M_{200}/[\hMsun])<12$ and $12<\newl ( M_{200}/[\hMsun]<13)$) is especially large at all radii and its radial dependence is significantly steeper than in all other cases.

The right panel of Fig.~\ref{fig:galtype_star_vs_halo_MinR_orient} shows the ratio, $\cos{\theta_{\rm star}^{\rm late}}/\cos{\theta_{\rm star}^{\rm early}}$, of the cosine of the misalignment angle between the stars of early- and late-type galaxies and the entire halo. At all radii of interest here, early-type galaxies are more misaligned than late-type galaxies. The misalignment  angle of late-type galaxies is smaller by about 10-20\% at $r\sim 0.03 r_{200}^{\rm crit}$ approximately the expected physical extent of the galaxy. 

A more detailed investigation of the galaxy-halo misalignment as a function of galaxy type is beyond the scope of this paper. We do acknowledge that this is certainly an interesting direction to be further explored, especially in view of the fact that many (current and forthcoming) lensing studies for which the misalignment angle hampers the interpretation of the signal use early-type galaxies such as LRGs. This exploratory work suggests that late-type galaxies are instead less misaligned with their host halo and therefore, in this respect, to be preferred to early-type galaxies.

\subsubsection{Misalignment for central and satellites galaxies}
\label{Sec:satcen_star_vs_halo_MinR_orient}

\begin{figure*} \begin{center} \begin{tabular}{cc}
\includegraphics[width=1.0\columnwidth]{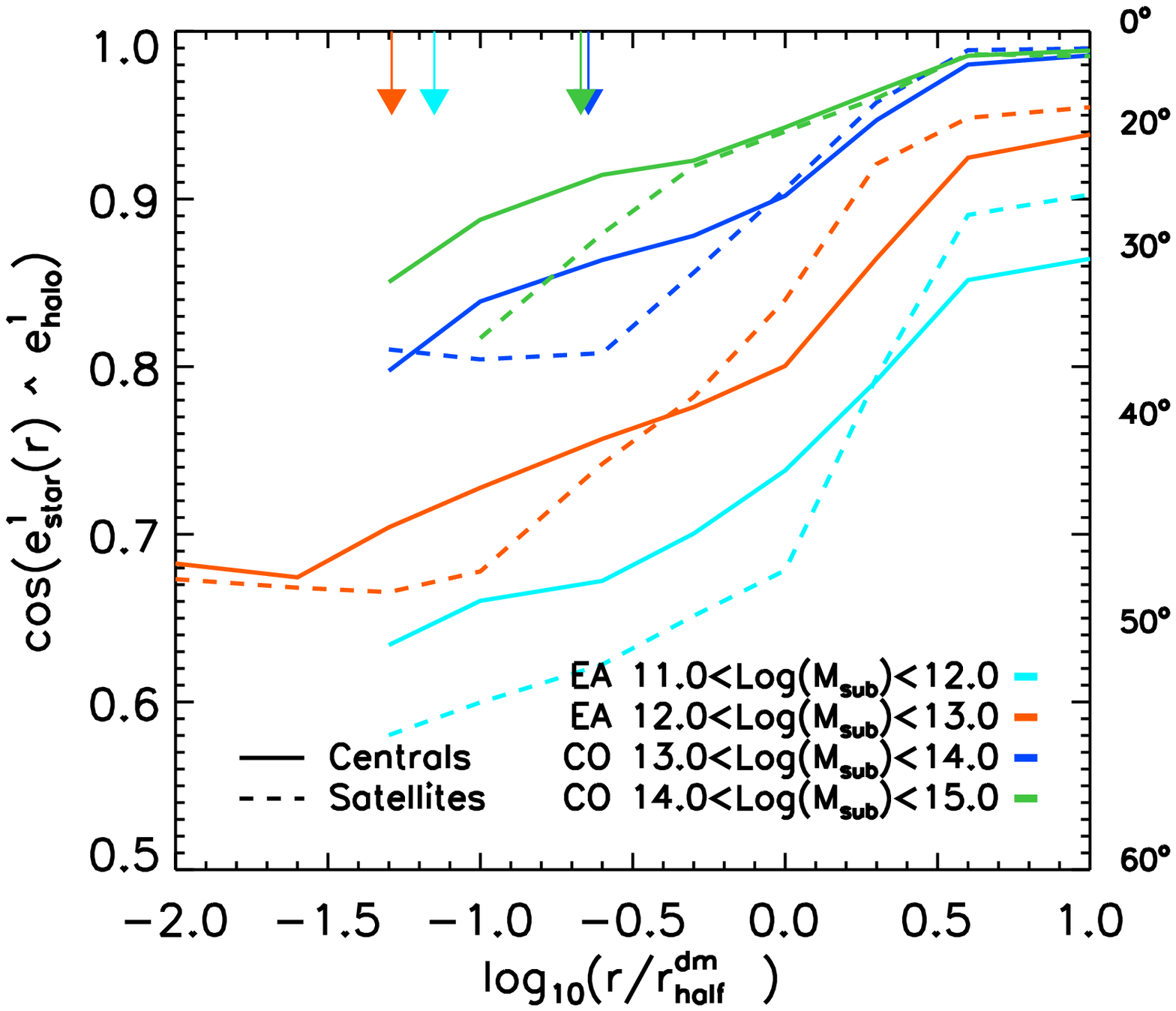} &
{\includegraphics[width=1.0\columnwidth]{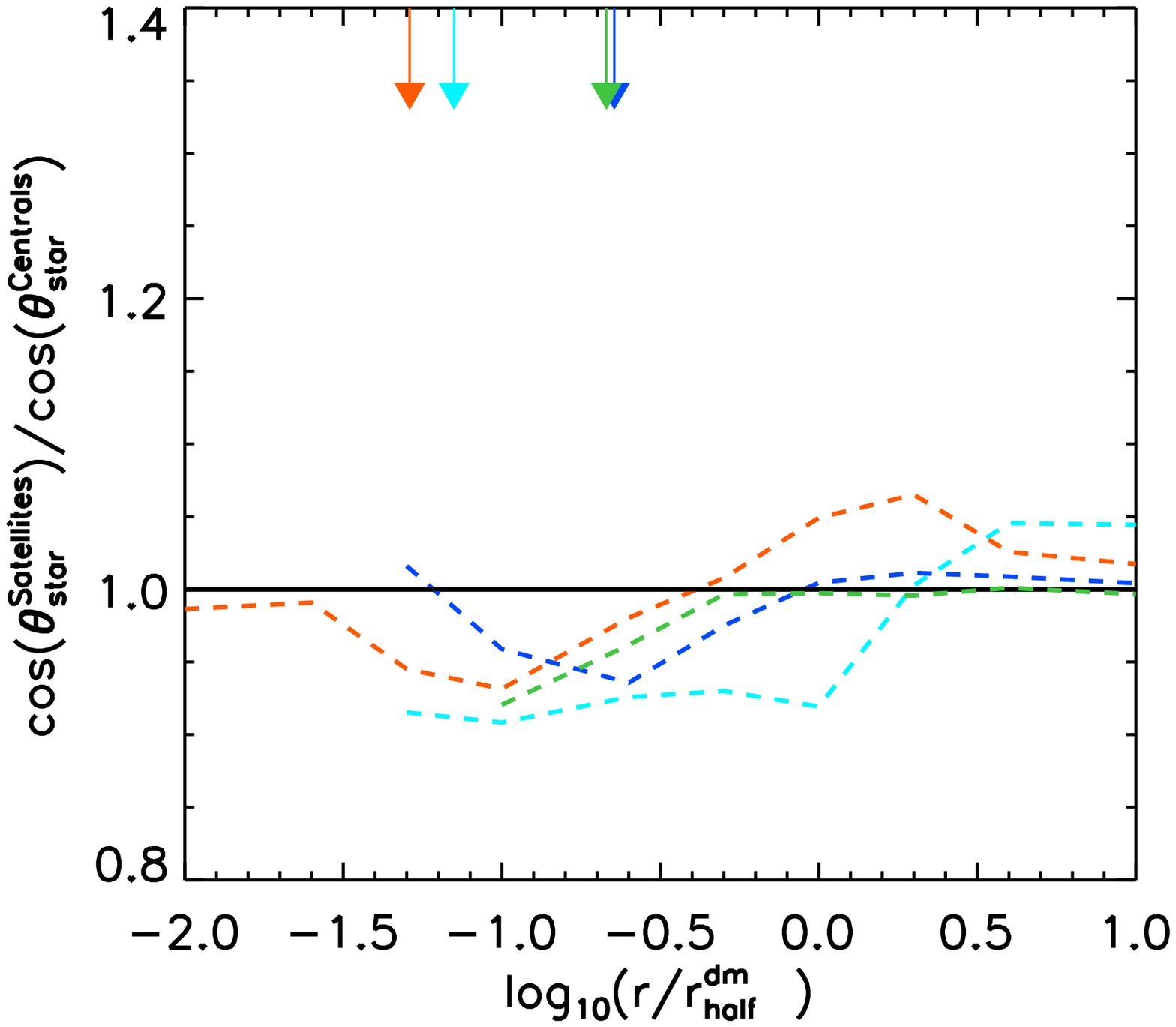}} \\
\end{tabular} \end{center}
\caption{
Same as Fig.~\ref{fig:galtype_star_vs_halo_MinR_orient} but for central and satellite galaxies and subhaloes (see text). To ease the comparison for results of different halo types, radii have been rescaled to the half mass radius for the dark matter mass, $r_{\rm half}^{\rm dm}$.
In both panels the least massive bin is taken from the L100 simulation in order to improve the statistic.
The vertical arrows represent the median value of $r_{\rm half}^{\rm star}$ in units of $r_{\rm half}^{\rm dm}$ in the different mass bins.
}\label{fig:satcen_star_vs_halo_MinR_orient}
\end{figure*} 

In this section, we characterize the alignment of stars with their host halo for centrals and satellites separately. Note that we have only considered centrals in the preceding sections. The left panel of Fig.~\ref{fig:satcen_star_vs_halo_MinR_orient} shows the median misalignment of the direction of the entire halo with that of the stars in central and satellite galaxies, whereas the right panel of Fig.~\ref{fig:satcen_star_vs_halo_MinR_orient} shows the ratio of the cosine of the misalignment angle between the entire halo and the stars for central and satellite galaxies. As for  Fig.~\ref{fig:galtype_star_vs_halo_MinR_orient}, we employ the EAGLE L100 simulation for the mass bins  ($11<\newl(M_{\rm sub}/[\hMsun])<12$ and $12<\newl(M_{\rm sub}/[\hMsun])<13$) to  improve the otherwise poor statistics of the EAGLE L025 simulations. Furthermore, we adopt here the dark matter half-mass radius, $r_{\rm half}^{\rm dm}$, as a definition of the extent of a halo, as this is properly defined for both centrals and satellites whereas an overdensity with respect to a background/critical value is an ill-defined concept for subhaloes that host satellite galaxies. At all radii, the misalignment angle between the entire halo and the stars in central and satellite galaxies is the same to within 10\%. The radial trend  is in qualitative agreement with those of the  whole sample shown in Fig.~\ref{fig:star_vs_halo_MinR_orient} (i.e. the misalignment decreases from the inner to the outer halo). 

The consistently lower misalignment in the outer parts of satellites could be due to the tidal stripping removing the outer (and more misaligned) part of the halo. Instead, in the inner part the resulting reduction of $r_{\rm half}^{\rm dm}$ (for which the radii are normalized)  would produce a shift of the whole relation to the right, effectively increasing the misalignment. The competition between these two processes could explain the transition between a more misaligned inner part to a less misaligned outer part of satellites with respect to central subhaloes.

This result indicates that satellite-specific physical processes (e.g. dynamical friction, tidal stripping) generally do not have a strong impact on the misalignment between the stellar and (mostly dark) matter component.

\subsubsection{The effect of projection on the misalignment angle}

\begin{figure*} \begin{center} \begin{tabular}{cc}
\includegraphics[width=1.0\columnwidth]{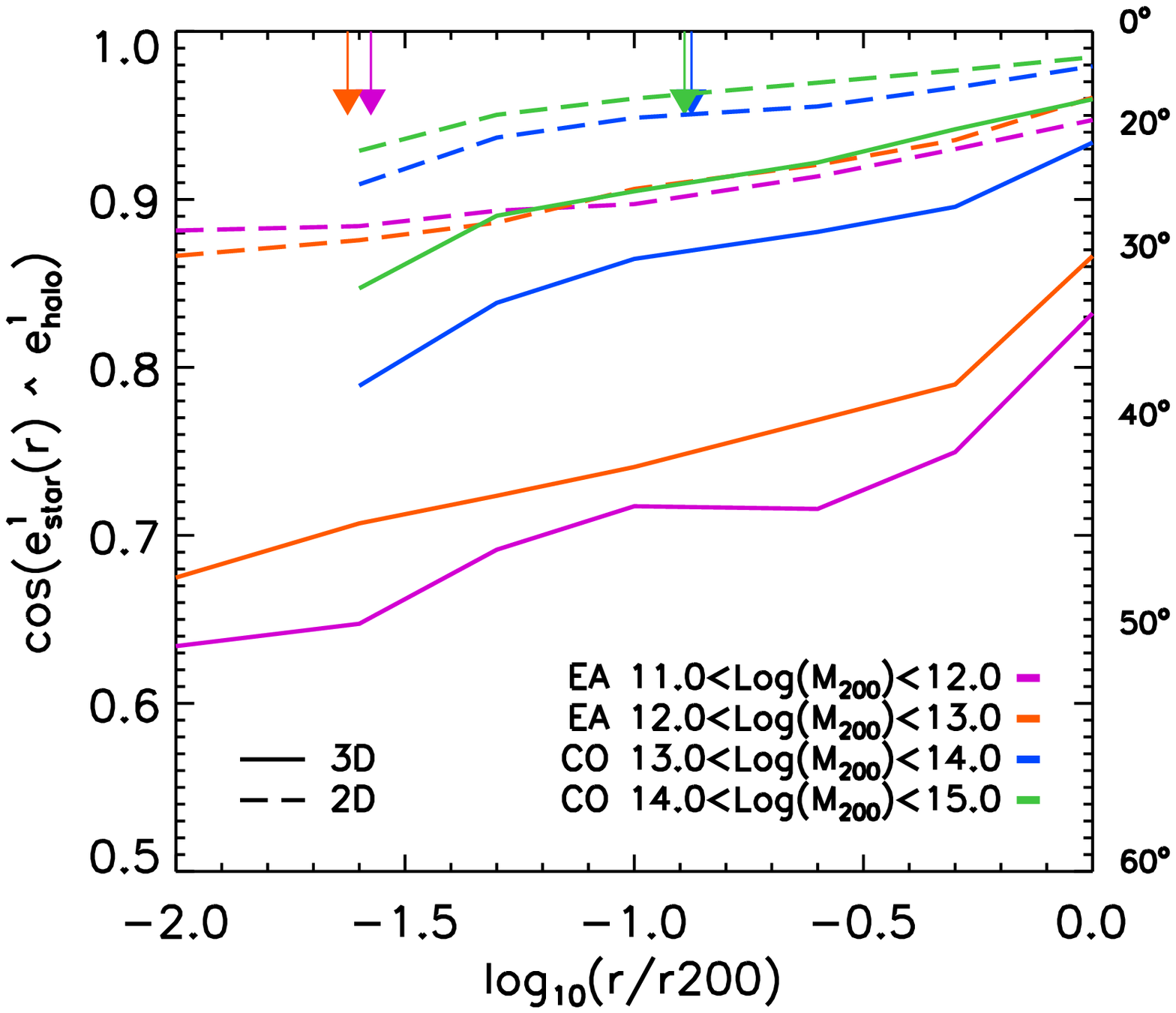} &
{\includegraphics[width=1.0\columnwidth]{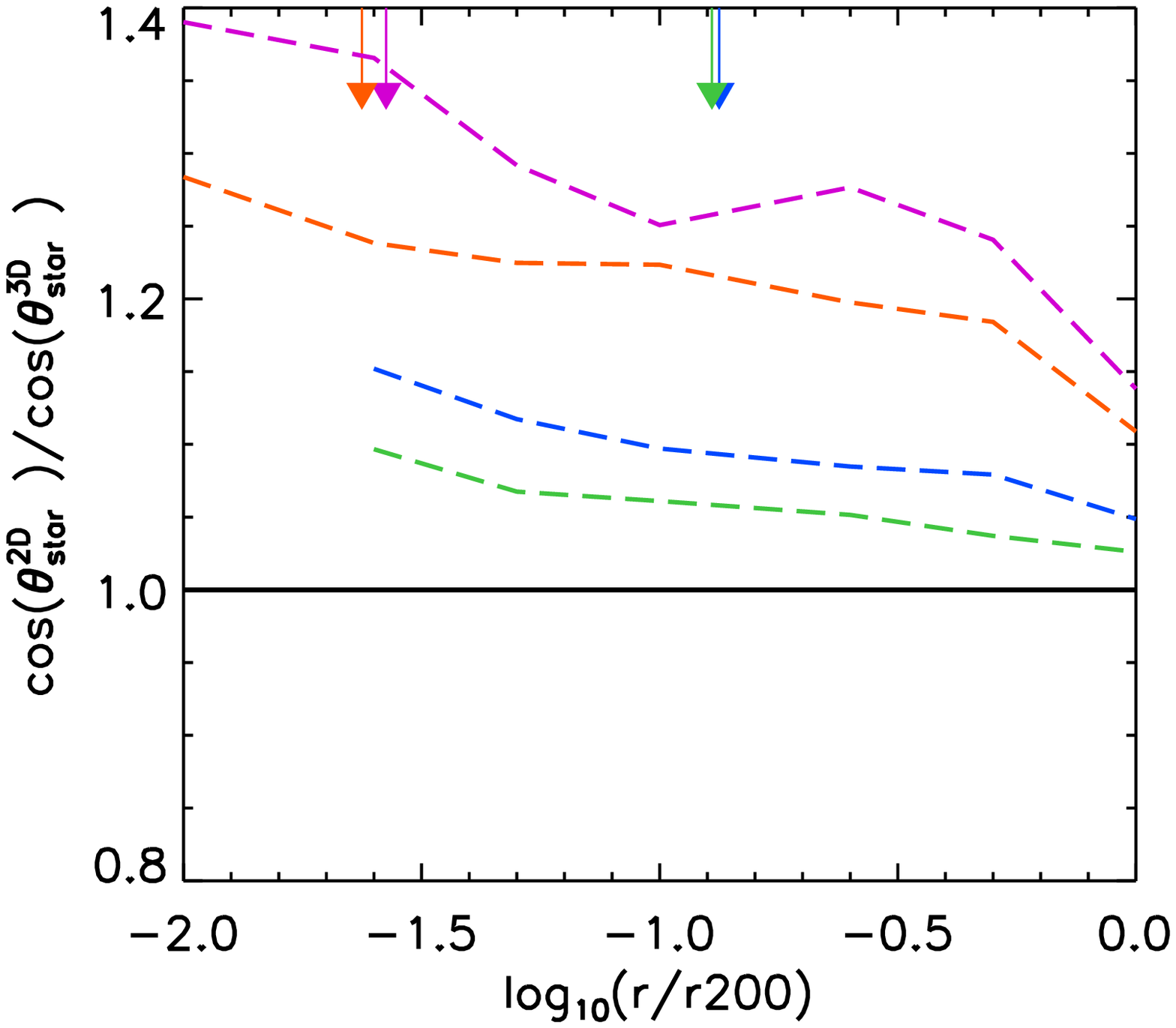}} \\
\end{tabular} \end{center}
\caption{
Comparison between the spatial variation of the 3D (continuous lines) and 2D (dashed lines) of the cosine of the median misalignment angles of stars with the underlying (mostly dark) matter distribution. Different colours refer to different halo mass bins. Only central haloes are used to exclude effects that can alter mostly the alignment of satellites. As discussed in the main body of the paper, the projection, by reducing the degrees of freedom of the system, increases the alignment.
}
\label{fig:2D3D_star_vs_halo_MinR_orient}
\end{figure*} 

Observationally one only has access to quantities projected onto the plane of the sky. Therefore, it is of interest to compute the misalignment in a (random) two-dimensional (2D) plane onto which all particles of the simulations have been projected. Correspondingly, one has 2D inertia tensors that describe the matter distribution of each component. In this 2D application, the misalignment angle between the stars and the halo is measured as the angle between the main eigenvectors of the inertia tensor of stars and (mostly dark) matter.

The left panel of  Fig.~\ref{fig:2D3D_star_vs_halo_MinR_orient} shows the radial- and mass-dependence of the median (cosine of the) misalignment angle for the 3D (solid) and the 2D (dashed) case.
Clearly, the net effect of projecting the 3D distribution onto a 2D plane is an increase in the alignment at all radii and all halo masses. 
The right panel of  Fig.~\ref{fig:2D3D_star_vs_halo_MinR_orient} shows the ratio between the cosine of the misalignment angle in 2D and 3D. The ratio decreases with both mass and radius  but is always greater than unity. It reaches values of about 1.25-1.35 for the low-mass bins at the radii that are representative of the physical extent of a galaxy. 
A similar result was reported in Ten14.

\subsection{Misalignment of hot gas with its host halo}
\label{sec:gasmisalignment}

\begin{figure*} \begin{center} \begin{tabular}{cc}
\includegraphics[width=1.0\columnwidth]{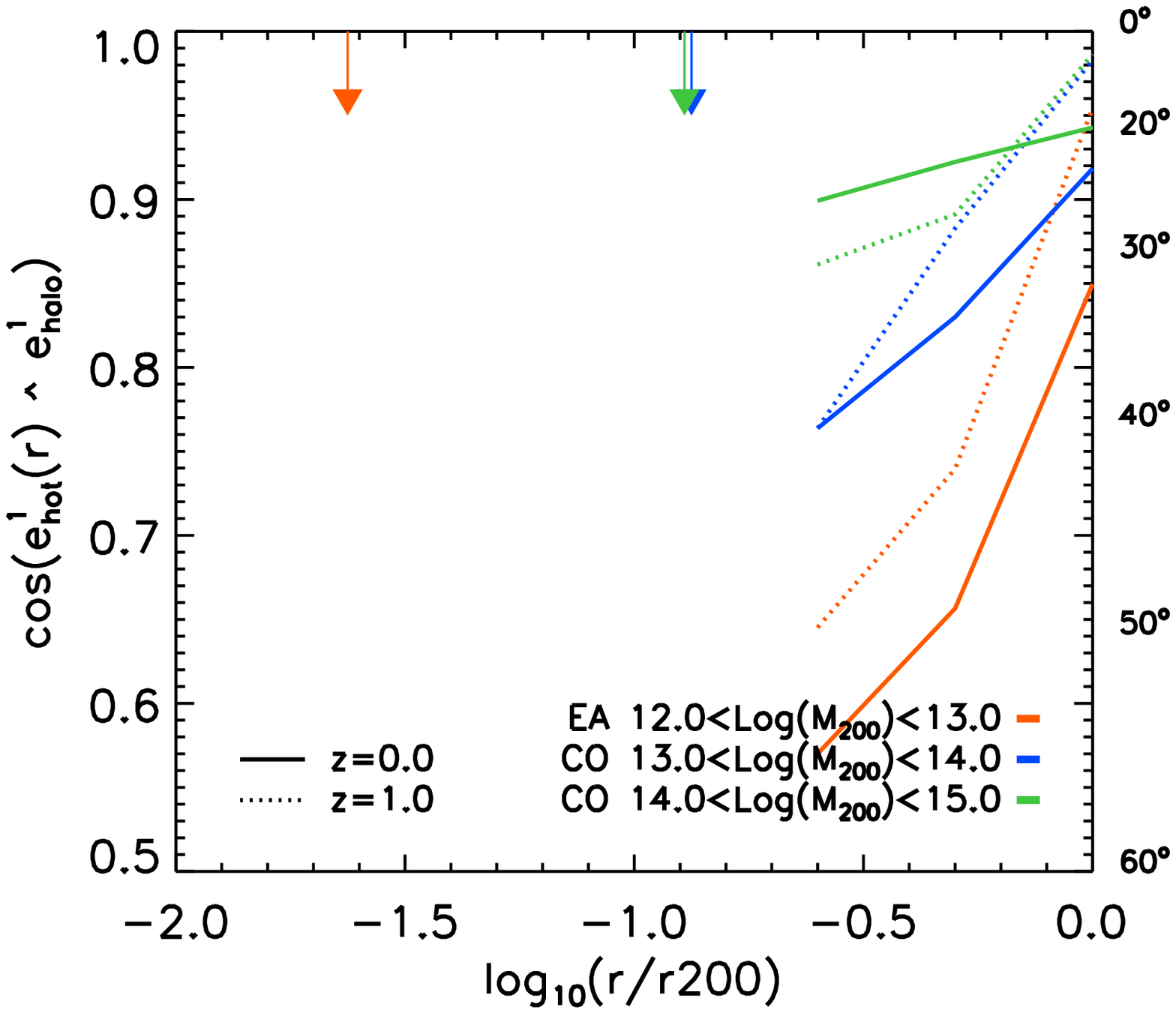} &
{\includegraphics[width=1.0\columnwidth]{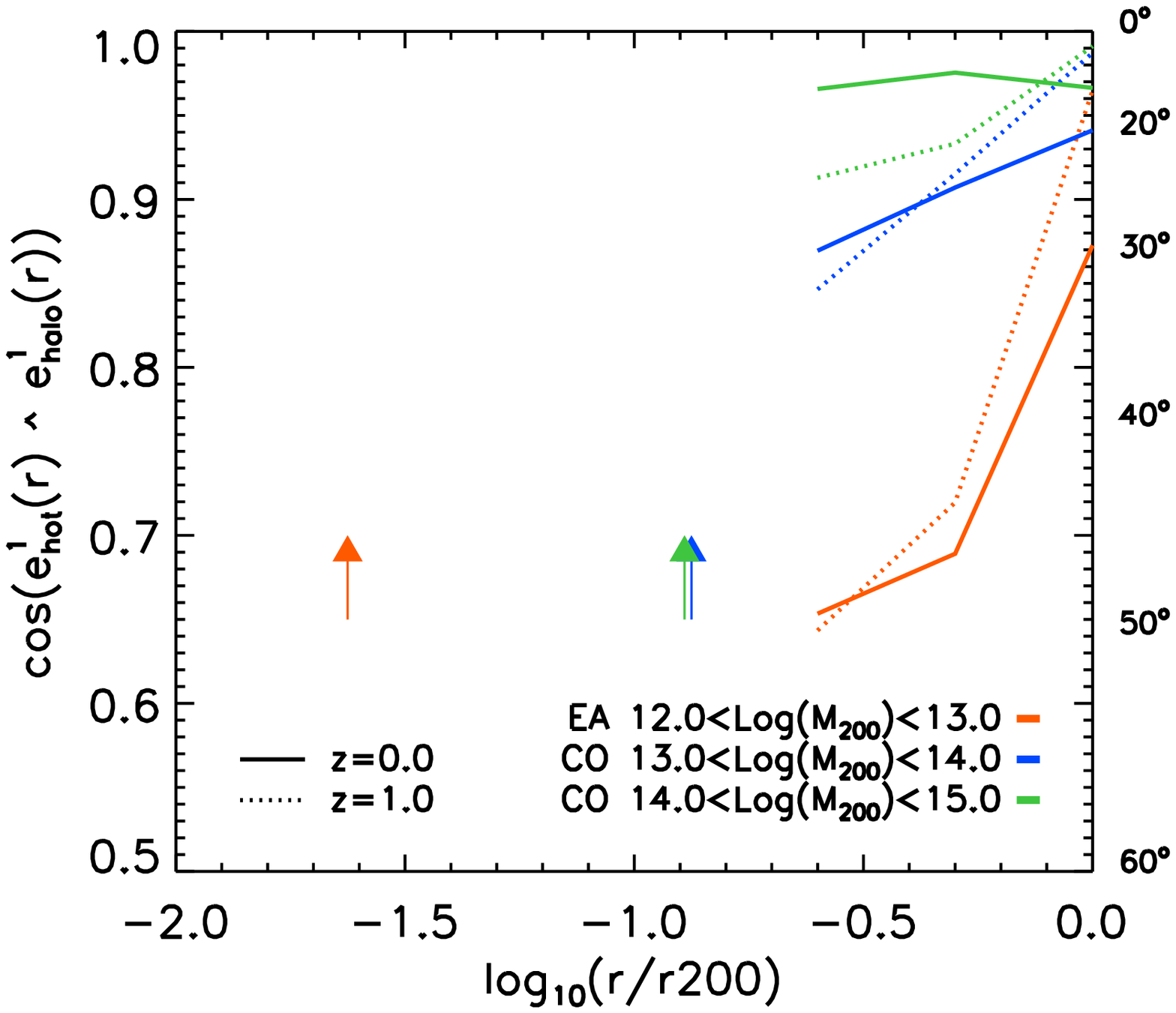}} \\
\end{tabular} \end{center}
\caption{
Same as Fig.~\ref{fig:star_vs_halo_MinR_orient} but for the hot ($T>10^6 \rm K$) gas component of haloes. The alignment of the hot component increases with radius and mass. Except for the highest-mass haloes, the gas  does not follow the dark matter distribution as well as was the case for the stars (c.f. Fig.~\ref{fig:star_vs_halo_MinR_orient}).
}
\label{fig:z0z1_hot_vs_halo_orient}
\end{figure*} 

Fig.~\ref{fig:z0z1_hot_vs_halo_orient}  shows the radial and mass dependence of the alignment of the hot component of the gas ($T>10^6 \rm K$) with its host halo. 
The results are only shown for three mass bins, because the mass bin $11< \newl(M_{200}/[\hMpc])< 12$ does not contain enough hot gas particles to retrieve reliable estimates for the orientation.
For the highest halo mass bins (right panel) the spatial variation of the misalignment angle between the hot component and the entire halo (left panel) is similar to that of the stars in the same halo mass bins. On the other hand, the misalignment between the hot gas and the local matter distribution differs from the case of stars: the hot gas component is significantly misaligned with respect to the local matter distribution. Specifically, for haloes in the mass range $12< \newl(M_{200}/[\hMsun]) < 13$ the  misalignment angle of the hot gas is as large as 50 degrees at $r\sim 0.3 \, r_{200}^{\rm crit}$ and it is $\sim$ 30 ($\sim$ 10) degrees for the halo mass range $13<\newl(M_{200}/[\hMsun])< 14$ ($14< \newl(M_{200}/[\hMsun])< 15$). 
Results for redshift $z=1$ (dotted lines) have similar radial and mass dependence as for redshift $z=0$.

Because it is observable out to larger radii than the stellar distribution of the central galaxy, hot gas represents a valuable tracer of the gravitational potential of massive clusters. Unfortunately, the fact that the hot gas tends to be largely misaligned with the local matter distribution makes it a poor tracer of the shape of the halo, unless $\newl(M_{200}/[\hMsun])> 14$.

\section{Summary and Conclusions}
\label{Sec:Conclusions}

This paper reports the results of a systematic study of halo and galaxy shapes and their relative alignment in the EAGLE \citep{Schaye14,Crain15} and cosmo-OWLS \citep{LeBrun14,McCarthy14} hydro-cosmological simulations.
Several aspects of these simulations make them an ideal tool for this investigation. First, the combination of these simulations allows us to apply our study to four orders of magnitude in halo masses with sufficient resolution and statistics. Second, the EAGLE simulations have been calibrated to be in agreement with the observed present-day galaxy stellar mass function and the observed size-mass relation (\citealt{Schaye14}). 
Third, it has been shown that cosmo-OWLS simulations reproduce key (X-ray and optical) observed properties of galaxy groups as well as the observed galaxy mass function for haloes more massive than $\log(M/[\hMsun]) = 13$.

We have studied the shapes of the distributions of dark matter, stars and hot gas in haloes with masses $11<\newl (M_{200}^{\rm crit}/[\hMsun])<15$
and their evolution in the redshift range $0 \le z \le 1$.
We find that the matter distribution in haloes is more aspherical (and triaxial) at higher halo mass and higher redshift (see Fig.~\ref{fig:ST_halo_vs_Msub}). The same qualitative trends hold for the star and the hot gas distribution in haloes (see Figs.~\ref{fig:ST_star_vs_Msub} and \ref{fig:ST_hot_vs_Msub}). We report (in Fig.~\ref{fig:star_STinR}) the spatial variation of the median of the shape parameters of the stellar distribution from $\sim0.02 r_{200}$ (i.e. a few to tens of kpc) to $r_{200}$ (i.e. up to a few Mpc). We note that at fixed radius and halo mass, stellar distributions are generally less spherical than dark matter haloes. We have measured the r.m.s. of the projected stellar ellipticity as a function of halo mass. We find a modest mass dependence, with r.m.s. stellar ellipticity increasing by 50 \% as halo mass increases by four orders of magnitude. We note that the values of the r.m.s. stellar ellipticity vary from $\sim0.2$ to $\sim0.35$ when one considers only stars within the star half-mass radius. However, the same quantity varies from $\sim0.35$ to $\sim0.55$ when all stars within the halo are considered (see Fig.~\ref{fig:erms_vs_Msub}).

\citet{Tenneti14a} recently used the Massive Black II simulation to study the mass dependence and evolution of the stellar and dark matter components of haloes and subhaloes. Their findings are, for the most part, in qualitative agreement with ours. However we find a few differences as reported in the corresponding sections (see e.g. \S~\ref{Sec:ShapeParametersResults} and the discussion of Fig.~\ref{fig:ST_star_vs_Msub}). Specifically, we highlighted sources of potential biases in their analysis. As detailed in \S~\ref{Sec:CaveatsGalForm}, those biases mostly stem from the use of a hydro-simulation that does not reproduce the observed stellar-halo mass relation and by imposing an artificial cut-off in the minimum stellar mass for which the shape is calculated.

We have measured the misalignment of the baryonic components (stars and hot gas) of galaxies with their own host haloes. We find that stars align well with the underlying (mostly dark) matter distribution, especially when all stars inside the halo are considered (see Fig.~\ref{fig:star_vs_halo_MinR_orient}). 
However, the stellar distributions in the inner parts of the host haloes do exhibit a median misalignment of about 45-50 degrees.
The misalignment is smaller in more massive haloes ($13 \le \newl(M_{200}/ [\hMsun]) \le 15$), late-type galaxies (see Fig.~\ref{fig:galtype_star_vs_halo_MinR_orient}), and central galaxies (see Fig.~\ref{fig:satcen_star_vs_halo_MinR_orient}). 
The hot gas distribution can only be traced with a sufficient number of particles only in the outer part ($\ge 0.3 r_{200}$) of massive ($12 \le \newl(M_{200}/ [\hMsun]) \le 15$) haloes. 
In this range we find that the alignment of the hot gas with the \emph{entire} halo is similar  to that of the stellar distribution. However, the hot gas does not align well with the \emph{local} matter distribution, exhibiting misalignment angles larger than 20 (typically 30 to 50) degrees in haloes with masses $13 \le \newl(M_{200}/ [\hMsun]) \le 15$ (see Fig.~\ref{fig:z0z1_hot_vs_halo_orient}). 

We have quantified the effect of projection on the median misalignment angles between the stellar distribution and the halo (see Fig.~\ref{fig:2D3D_star_vs_halo_MinR_orient}). Projection reduces the degrees of freedom of the system, increasing the alignment.  Finally, we provided the probability distribution of the misalignment angle between the major axis of the stellar distribution inside the stellar half-mass radius and the major axis of the entire halo for the three- and two-dimensional case (see Figs.~\ref{fig:star_inhalo_orient_histo} and \ref{fig:star_inhalo_orient_histo_2d}, respectively).

We have encapsulated our results in fitting functions (see Appendix B) and tables that allow interested practitioners to straightforwardly include our results into halo catalogues extracted from N-body simulations. The complete list of fitting parameters as well as tabulated values are available at http://www.strw.leidenuniv.nl/MV15a/.

A natural extension of this work is the study of the correlation functions of galaxy shapes. 
We will present such an investigation in a future publication.


\section*{Acknowledgements}  
\label{sec:acknowledgements}
We thank the anonymous referee for insightful comments that helped improve the manuscript. MV and MC thank Henk Hoekstra and Rachel Mandelbaum for useful and stimulating discussions.  MC acknowledges support from NWO VIDI grant number 639.042.814
and ERC FP7 grant 278594. RAC is a Royal Society University Research Fellow. This work used the DiRAC Data Centric system at Durham University, operated by the Institute for Computational Cosmology on behalf of the STFC DiRAC HPC Facility (www.dirac.ac.uk). This equipment was funded by BIS National  E-infrastructure capital grant ST/K00042X/1, STFC capital grant ST/H008519/1, and STFC DiRAC Operations grant ST/K003267/1 and Durham University. DiRAC is part of the National E-Infrastructure. We also gratefully acknowledge PRACE for awarding us access to the resource Curie based in France at Tr\`es Grand Centre de Calcul. This work was sponsored by the Dutch National Computing Facilities
Foundation (NCF) for the use of supercomputer facilities, with financial support from the Netherlands Organization for Scientific
Research (NWO). The research was supported in part by the European Research Council under the European Union's Seventh Framework
Programme (FP7/2007-2013) / ERC Grant agreements 278594-GasAroundGalaxies, GA 267291 Cosmiway, and 321334 dustygal, the UK Science and Technology Facilities
Council (grant numbers ST/F001166/1 and ST/I000976/1), Rolling and Consolodating Grants to the ICC, Marie Curie Reintegration Grant PERG06-GA-2009-256573. TT acknowledge the Interuniversity Attraction Poles Programme initiated by the Belgian Science Policy Office ([AP P7/08 CHARM])


\bibliographystyle{mn2e}
\bibliography{paper} 
\expandafter\ifx\csname natexlab\endcsname\relax\def\natexlab#1{#1}\fi

\appendix
\section{Caveats in shape parameter estimation}
\label{Sec:caveats}

\subsection{The choice of inertia tensor}

There exists a plethora of methods designed to characterize the shape of a given three-dimensional particle distribution (say dark matter, star, gas) 
in the context of cosmological structure formation simulations (see Zemp et al. 2011 and references therein). All those methods are based on the idea that structures can be well described by an ellipsoidal shape. However, the actual algorithms used to retrieve this shape can differ substantially 
and unfortunately the corresponding results do not often agree (see Zemp et al. 2011 for an analysis of this problem under controlled conditions with 
known shapes). Most notably, results on the shape of a particle distribution may vary if one adopts the inertia tensor rather than the reduced inertia 
tensor, or some iterative form of the two \citep[see discussions in][]{Zemp11, Tenneti14b}. 
The differences between the inertia tensor and the reduced inertia tensor are driven by the fact that in the reduced inertia tensor calculation particles are not weighted by their distance from the centre.
The net effect is that if the reduced inertia tensor is used, the shape is less dominated by the particles in the outer part of haloes, meaning that the retrieved shape tends to be more spherical as particles in the inner parts of haloes are typically more spherically distributed.
We repeated our analysis using the reduced inertia tensor and found that there is little information content in exploring the radial variation using this method since its properties have almost no variation with radius.
We also note that the misalignment angle, which is the quantity of primary interest here, is less affected than shape by the choice of the algorithm that defines the shape parameters. This is especially true when the alignment is calculated between particles distributions at the same distance from the centre.

Another possible variation in the shape calculation is to use an iterative method for the inertia tensor calculation but this method was proven to give very similar results when the inertia tensor is used, as shown in \citet{Tenneti14b}.

Throughout the paper we adopted the definition presented in 
eq.~(\ref{eq:inertiatensor}) and the corresponding shape parameters defined in eq.~(\ref{eq:shapeparameters}). While our adopted method may be considered somewhat arbitrary \citep[e.g. see the discussions in][]{Jing02, Zemp11}, we used this approach as it is adequate 
for the comparison we presented \citep[see e.g][]{Bett12} and because it allowed  us to compare our results with most of the other results in the literature.

\subsection{The effect of sampling}
\label{Sec:ShapeSampling}
An important technical aspect regarding shape measurements is to find the minimum number of particles required to obtain a reliable estimate. 
To this aim, we simulate a three-dimensional halo with a given axis ratio and a Navarro, Frenk, \& White (hereafter NFW) density profile \citep{NFW}. 
More specifically, we choose values for the three-dimensional halo axis a, b and c, and use an analyitical NFW profile with $c=5$ and $r_{\rm vir}=a$ (the largest axis).
We then generate $N_{\rm part}$ spherical coordinates $(r,\phi,\theta)$ using the NFW profile as a selection function, redrawing any coordinates that fall outside the ellipsoid defined by a, b, and c.
Specifically, we use $1 \le N_{\rm part} \le 3000$. For each value of $N_{\rm part}$, we repeat the sampling $10^5$ times so as to obtain a median and a standard deviation. 

It is worth noting that 
the number of particles needed for an unbiased shape determination depend on the intrinsic shape of the halo. Many more particles are needed to retrieve a quasi-spherical shape than for example a disky structure. For our test, the intrinsic shape of the halo was chosen to have sphericity $S=0.6$ and triaxiality $T=0.7$, which is representative of the average shape parameters of our halo sample (see e.g. results in \S\ref{Sec:star_STinR} and Fig.~\ref{fig:star_STinR}).

\begin{figure} 
\begin{center} 
\includegraphics[width=1.0\columnwidth]{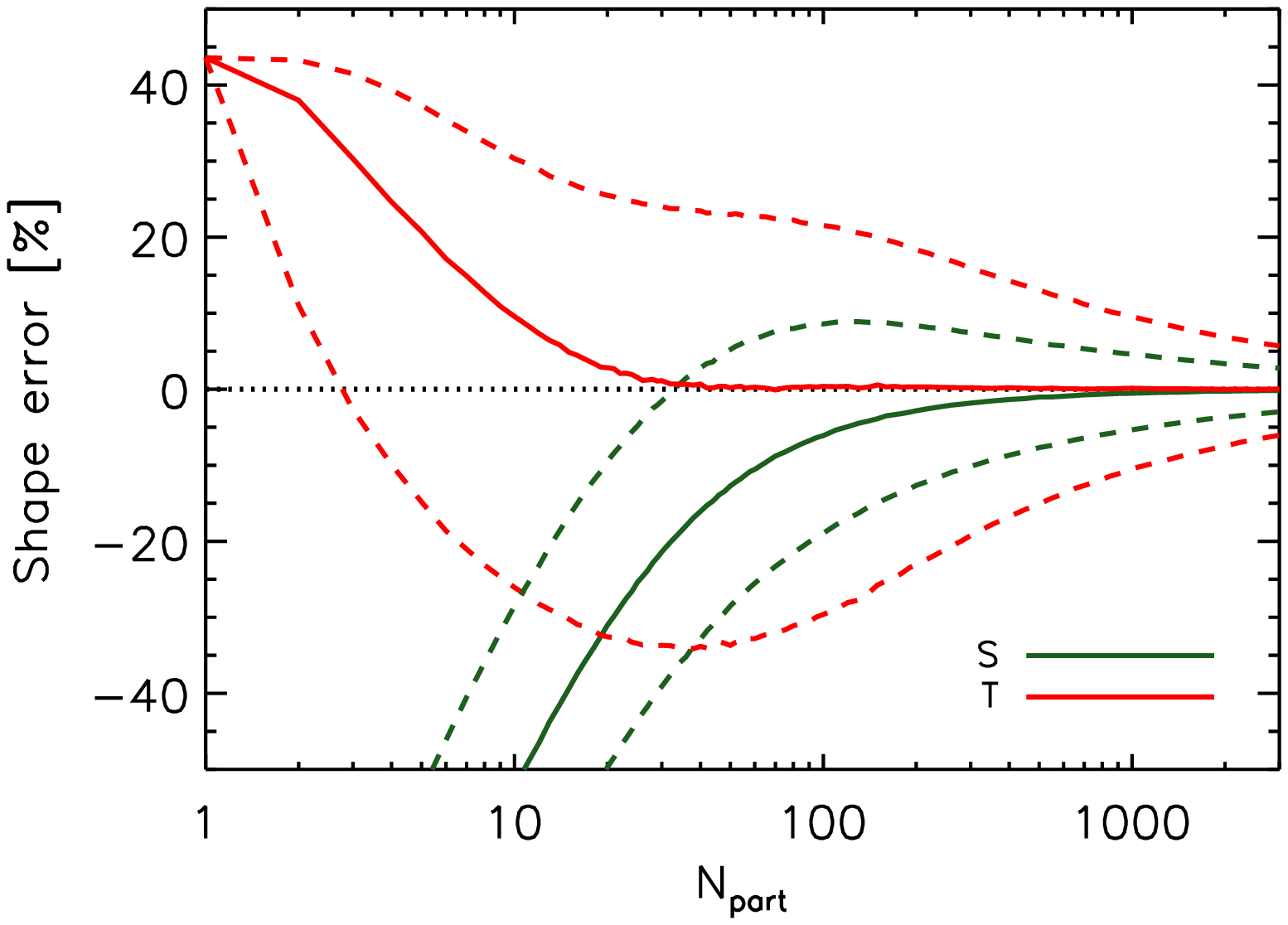}
\end{center}
\caption{
Convergence test for shape parameter retrieval. Relative error on the retrieved shape parameter of a synthetic NFW halo as a function of the number of particles used to sample the underlying distribution. The relative errors on sphericity and triaxiality are indicated by the green and red lines, respectively. The test is performed using a typical sphericity value for the synthetic halo, $S=0.6$ and $T=0.7$. For each number of particles, distributions are drawn $10^5$ times and we report the 50th (continous lines), 16th and 84th percentiles (dashed lines). Retrieving both shape parameters with a systematic error smaller than a few percent requires at least 300 particles.
}
\label{fig:shapetest}
\end{figure}

In Fig.~\ref{fig:shapetest} we show the relative error on the retrieved shape parameters, $S$ (green lines) and $T$ (red lines), as a function of $N_{\rm part}$. Solid lines refer to the median, whereas dashed lines refer to the $16 \rm th$ and $84 \rm th$ percentiles. 
The retrieved sphericity shows a monotonic trend with the number of test particles. The sphericity increases towards the real value as the number of test particles is increased. This means that any resolution effect will lead to an underestimating of the true sphericity of haloes. For this particular halo shape using 300 particles will lead to an average $\sim 2\%$ error in the determination of the sphericity with an accuracy of $\sim 10\%$.
The triaxiality is typically underestimated but converges faster to the true value, with the systematic error dropping below 3\% for 30 particles. On the other hand the scatter around the median converging slowly and is still $20\%$ for 300 particles.
Triaxiality thus requires more particles than sphericity in order to reduce the random error below a specific value. 
Throughout the paper, we thus employ $N_{\rm parts}\ge 300$ as the limit for shape parameter determination. This assures very good estimate of the median value of the shape parameters with a systematic error below $3\%$ and a random error of $10\%$ in the sphericity and $20\%$ in triaxiality. In this work we did not show these systematic errors in the shape measurement but only the statistical errors evaluated using the bootstrapping technique.

\section{Analytic fits for the misalignment angle distributions}
\label{Sec:fits}
In this section we provide fitting functions\footnote{The analytic fits provided in this section reproduce the median of the distributions obtained from the simulations  with an accuracy better than 1\%.} for the distribution of the cosine of the 3D misalignment angle $\theta$, as well as for the 2D misalignment angle, $\Theta$. We note that the choice of using the cosine as the variable of the fitting function stems from the notion that the distribution of the cosine of the alignment angle of a random set of 3D vectors is flat, whereas the distribution of the angle itself is not, as it is skewed towards large alignments.

We employ the following functional form:
\begin{equation}
{\cal M}^{\rm 3D}(x) = A + \exp{[B -\lambda (1 -x)^{\beta}]} \, ,
\label{eq:star_inhalo_orient_fit}
\end{equation}
where $x = \cos{(\theta)}$ and $0<\theta<\pi/2$. This functional form has four free parameters: $A, B, \lambda, \beta$. 
We find this number of parameters necessary to adequately reproduce the main features of the results obtained from the simulations. 
In the main body of the paper (see Fig.~\ref{fig:star_inhalo_orient_histo}),
we have employed this fitting function to describe the misalignment angle between the stellar component and its host halo in four halo mass bins and for the typical extent of a galaxy, the half mass radius $r^{\rm star}_{\rm half}$. 
The corresponding fitting parameters are given in Table~\ref{tbl:star_inhalo_orient_fit}. Parameters that refer to other components, radius definitions, and halo mass bins, as well as tabulated median values, can be found at http://www.strw.leidenuniv.nl/MV15a/.

We analytically describe the probability function of the cosine of the 2D misalignment angle with the following functional form:
\begin{equation}
{\cal M}^{\rm 2D}(x) = C \exp \left( - \frac{x^2}{2\sigma_1^2} \right) + D \exp\left( - \frac{x^2}{2\sigma_2^2} \right) + E \, ,
\label{eq:star_inhalo_orient_fit_2d}
\end{equation}
where $C, \sigma_1, D, \sigma_2, E$ are the 5 free parameters required to describe a double Gaussian plus a `floor'. The level of complexity of this functional form is motivated by the results obtained from the simulations. In the main body of the text (see especially Fig.\ref{fig:star_inhalo_orient_histo_2d}),
we describe the probability distribution of the 2D misalignment angle between stars and their host haloes in four halo mass bins and accounting only for stars within the  typical extent of a galaxy, the half mass radius $r^{\rm star}_{\rm half}$. The corresponding fitting parameters are given in Table~\ref{tbl:star_inhalo_orient_fit_2d}. Parameters that refer to other components, radius definitions, and halo mass bins, as well as tabulated median values, can be found at http://www.strw.leidenuniv.nl/MV15a/. It is instructive to compare these 2D misalignment angle distributions 
to the commonly assumed single-Gaussian distribution \citep[see e.g.][]{Okumura09}.
None of the distributions found in this study resembles a single-Gaussian and we therefore caution interested practitioners against adopting this assumption.  

\begin{table} 
\begin{center}
\begin{tabular}{lccccc}
\hline
Simulation &  mass bin & $A$ & $B$ & $\lambda$ & $\beta$ \\                           
\hline   
EA L025 &  $[11-12]$ & $1.52E-02$ & $-3.58$ & $5.92$ & $1.01$ \\
EA L100 &  $[12-13]$ & $6.43E-03$ & $-0.05$ & $5.13$ & $0.15$ \\
CO L200 &  $[13-14]$ & $4.46E-03$ & $-1.04$ & $5.64$ & $0.41$ \\
CO L400 &  $[14-15]$ & $4.13E-03$ & $-0.53$ & $7.13$ & $0.42$ \\ 
\hline 
\end{tabular}
\caption{
Fit parameters for Eq.~\ref{eq:star_inhalo_orient_fit} that describes the misalignment angle distribution between the direction of the stellar component inside $r_{\rm half}^{\rm star}$ and that of the entire halo.} 
\label{tbl:star_inhalo_orient_fit} 
\end{center}
\end{table}

\begin{table} 
\begin{center}
\begin{tabular}{lccccc}
\hline 
mass bin & $\sigma_1$ & $\sigma_2$ & $C$ & $D$ & $E$  \\            
\hline                                                                          
$[11-12]$ & $5.00$ & $28.17$ & $4.69E-02$ & $4.69E-02$ & $4.69E-02$ \\
$[12-13]$ & $5.00$ & $31.65$ & $1.31E-02$ & $1.31E-02$ & $1.31E-02$ \\
$[13-14]$ & $14.70$ & $32.52$ & $3.61E-02$ & $3.61E-02$ & $3.61E-02$ \\
$[14-15]$ & $9.42$ & $25.71$ & $4.28E-02$ & $4.28E-02$ & $4.28E-02$ \\
\hline 
\end{tabular}
\caption{Fit parameters for the double Gaussian fitting function Eq.~\ref{eq:star_inhalo_orient_fit_2d} that describes the misalignment angle distribution between the direction of the projected stellar component inside $r_{\rm half}^{\rm star}$ and that of the entire (projected) halo.
} 
\label{tbl:star_inhalo_orient_fit_2d} 
\end{center}
\end{table}

\section{Resolution test}\label{Sec:res_test}
In this section we make use of our different simulations to test the influence of resolution on our results.
For this test we make use of the fact that the L025 simulation is the high-resolution version of L100 simulated using a smaller box size. The same is true for L200 and L400. We do not compare results from simulations that were not run with the same code.

\begin{figure*} \begin{center} \begin{tabular}{cc}
\includegraphics[width=1.0\columnwidth]{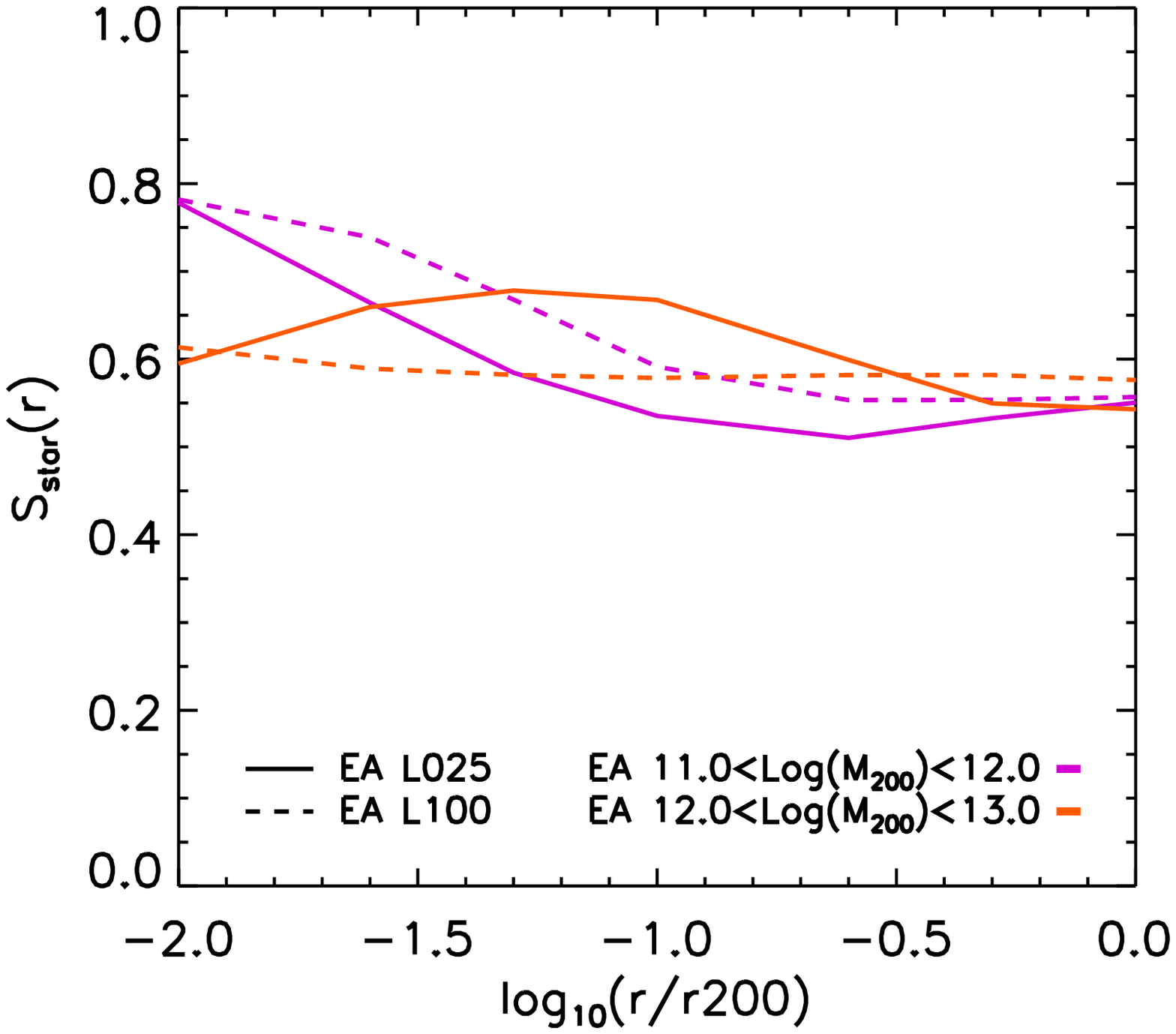} &
\includegraphics[width=1.0\columnwidth]{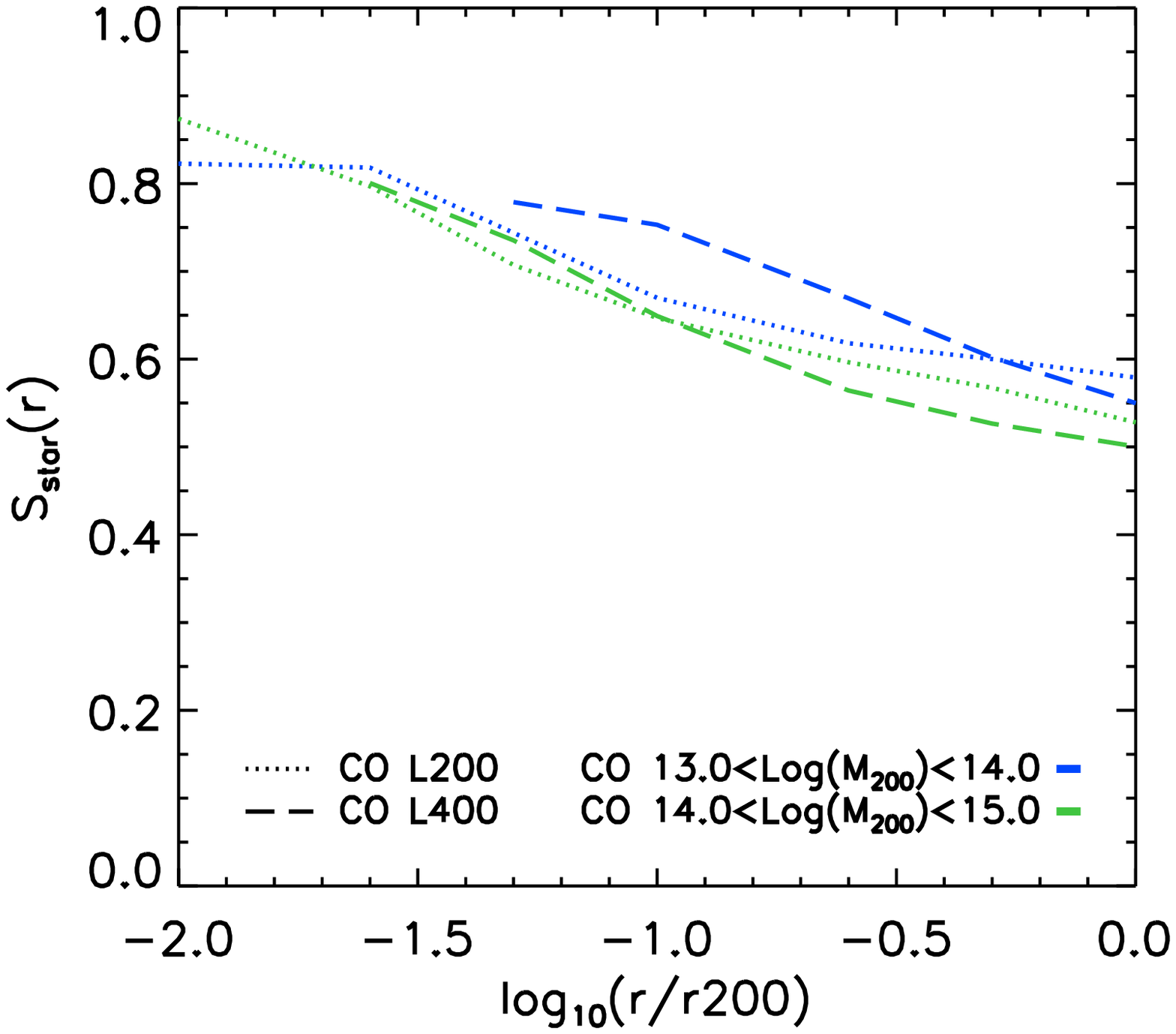} \\
\includegraphics[width=1.0\columnwidth]{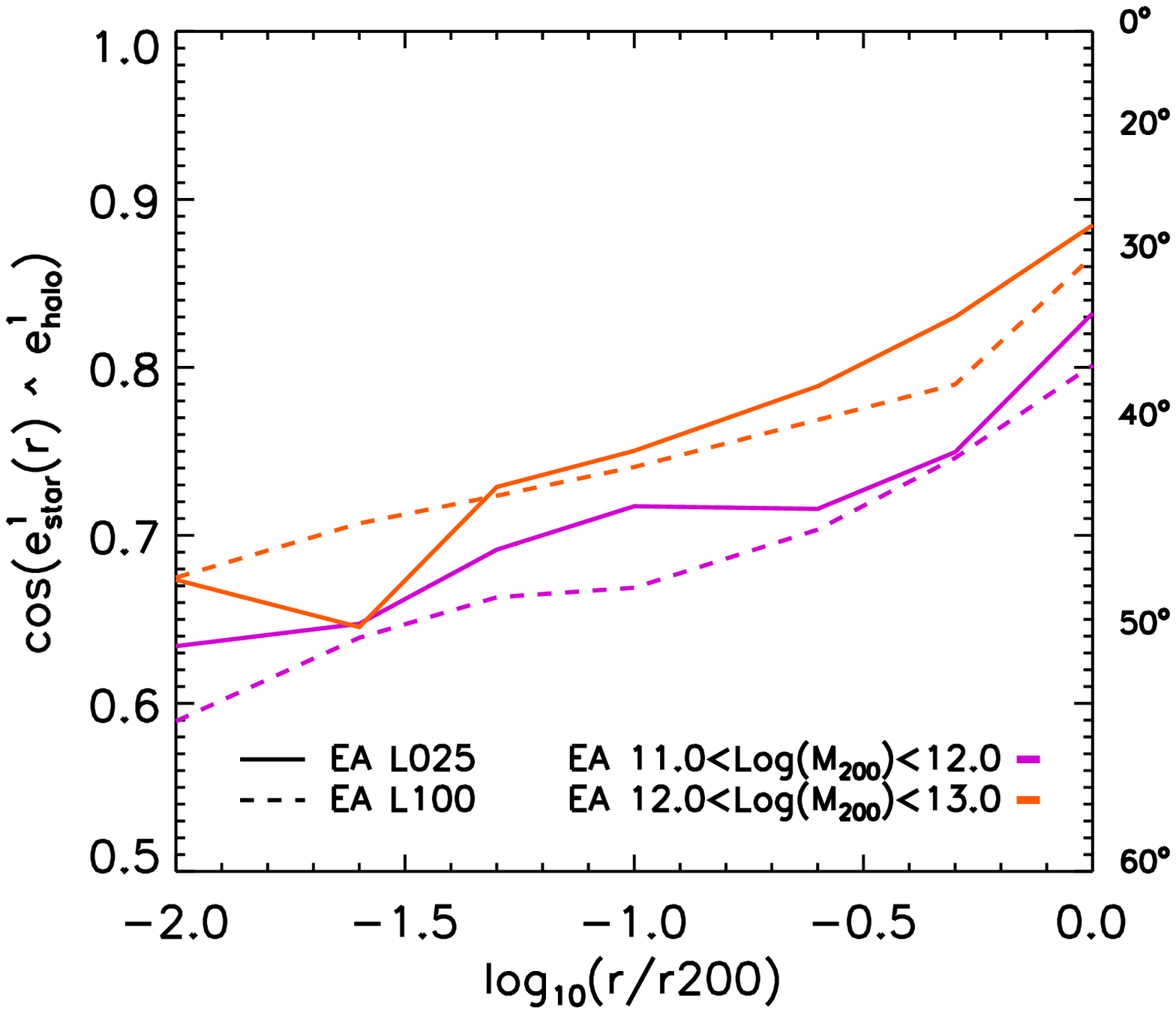} &
\includegraphics[width=1.0\columnwidth]{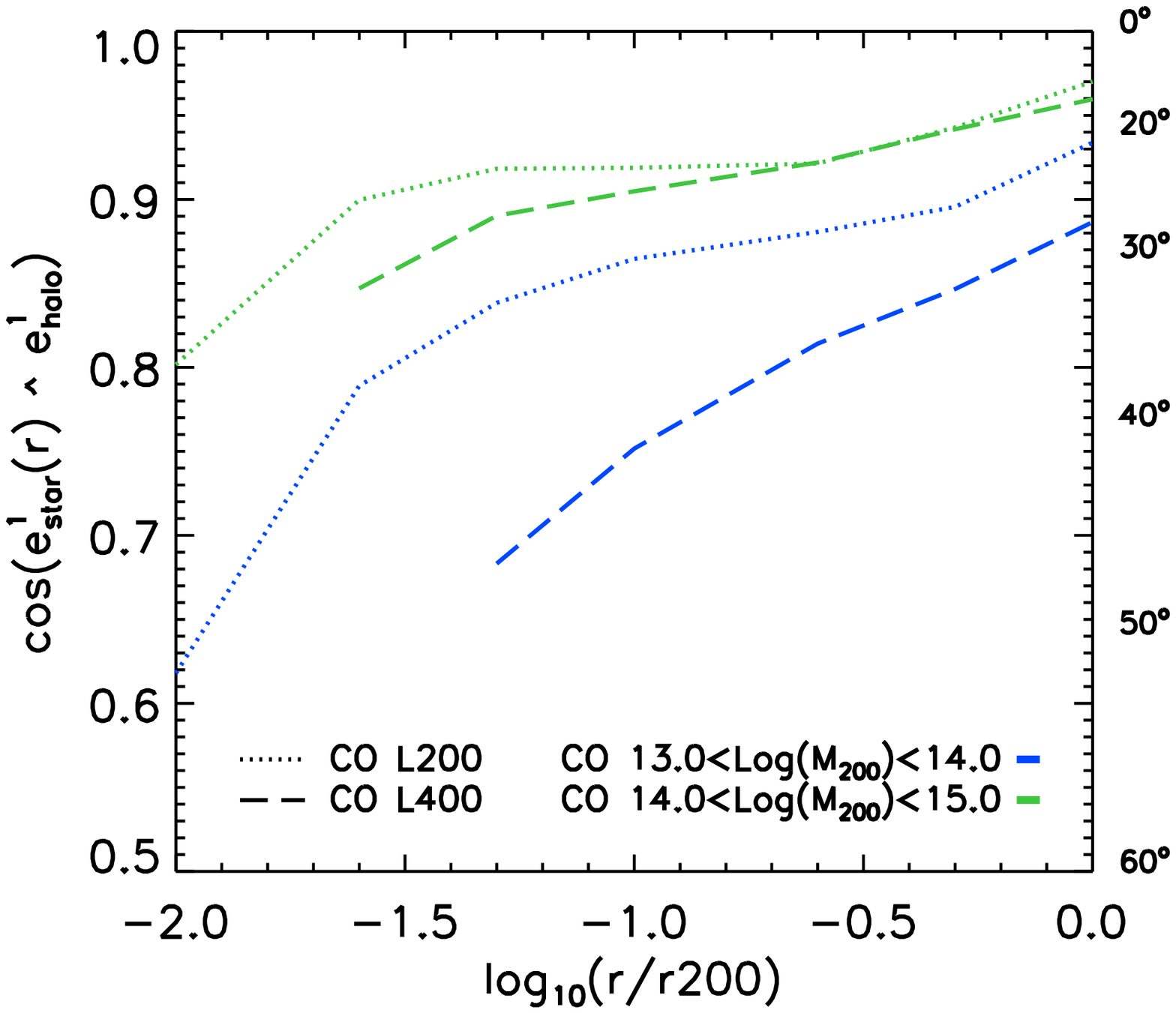} \\\end{tabular} \end{center}
\caption{ Resolution test for the variation of the sphericity (upper panels) and misalignment with the halo (lower panels) of the stellar component as a function of radius. We show separately the results for EAGLE (on the left) and cosmo-OWLS (on the right). For each set of simulations we show the results in two distinct mass bins. }
\label{fig:restest_orientation}
\end{figure*}

In Fig.~\ref{fig:restest_orientation} we show in the upper panels the variation of the sphericity of the stellar component. In the left panel we show the two mass bins for which is it possible to obtain results for both the L025 and L100 EAGLE simulations. On the right we do the same for the L200 and L400 cosmo-OWLS simulations. Different colours refer to different mass bins where as different line styles refer to different simulations.
In the lower panels we show in the same manner the misalignment between the stellar component and the whole halo.

The convergence is generally good, especially at larger radii, even though the box size, and hence the halo samples, also change between the different simulations. The only case that shows a relatively poor convergence is the misalignment for the least massive bin of the cosmo-OWLS simulations (blue lines) for which the shape of the curves are similar but the values are shifted between the two simulations.

\label{lastpage}
\end{document}